\documentclass[letterpaper,twocolumn,10pt]{article}
\usepackage{usenix2019_v3}

\usepackage{tikz}
\usepackage{amsmath}

\usepackage{cite}
\usepackage{amsmath,amssymb,amsfonts}
\usepackage{algorithmic}
\usepackage{graphicx}
\usepackage{textcomp}
\usepackage{xcolor}
\usepackage{makecell}
\usepackage{subfigure} 
\usepackage{url} 
\usepackage{booktabs,multirow,tabularx}
\usepackage{appendix}
\usepackage{paralist}
\usepackage{pifont}
\usepackage{enumitem}

\newtheorem{definition}{Definition}
\newtheorem{lemma}{Lemma}
\newtheorem{theorem}{Theorem}
\definecolor{mycolor}{HTML}{F0A12C}

\begin{document}
\date{}

\title{\Large \bf Fluent: Round-efficient Secure Aggregation for Private Federated Learning   
   \\
}

\author{
{\rm Xincheng Li}\\
Yangzhou University \\ 
n2308567e@e.ntu.edu.sg
\and
{\rm Jianting Ning}\\
Fujian Normal University \& \\
City University of Macau \\
jtning88@gmail.com
\and
{\rm Geong Sen Poh}\\
S-Lab for Advanced Intelligence \\
Nanyang Technological University \\
geongsen.poh@ntu.edu.sg
\and
{\rm Leo Yu Zhang}\\
Griffith University \\
leo.zhang@griffith.edu.au
\and
{\rm Xinchun Yin}\\
Yangzhou University  \\
xcyin@yzu.edu.cn
\and
{\rm Tianwei Zhang}\\
Nanyang Technological University \\
tianwei.zhang@ntu.edu.sg
}

\maketitle

\begin{abstract}
Federated learning (FL) facilitates collaborative training of machine learning models among a large number of clients while safeguarding the privacy of their local datasets. However, FL remains susceptible to vulnerabilities such as privacy inference and inversion attacks. Single-server secure aggregation schemes were proposed to address these threats. Nonetheless, they encounter practical constraints due to their round and communication complexities. This work introduces Fluent, a round and communication-efficient secure aggregation scheme for private FL. Fluent has several improvements compared to state-of-the-art solutions like Bell et al. (CCS 2020) and Ma et al. (SP 2023): (1) it eliminates frequent handshakes and secret sharing operations by efficiently reusing the shares across multiple training iterations without leaking any private information; (2) it accomplishes both the consistency check and gradient unmasking in one logical step, thereby reducing another round of communication. With these innovations, Fluent achieves the fewest communication rounds (i.e., two in the collection phase) in the malicious server setting, in contrast to at least three rounds in existing schemes. This significantly minimizes the latency for geographically distributed clients; (3) Fluent also introduces Fluent-Dynamic with a participant selection algorithm and an alternative secret sharing scheme. This can facilitate dynamic client joining and enhance the system flexibility and scalability. We implemented Fluent and compared it with existing solutions. Experimental results show that Fluent improves the computational cost by at least 75\% and communication overhead by at least 25\% for normal clients. Fluent also reduces the communication overhead for the server at the expense of a marginal increase in computational cost.
\end{abstract}

\section{Introduction}
Federated learning (FL)~\cite{McMahanAISTATS2017} has gained increasing attention from both academia and industry owing to its capability of learning from large amounts of geographically distributed data. In FL, each participating device (i.e., client) individually updates the weights of a machine learning model with a prescribed training algorithm on its local data. All the updated weights are then sent to a coordinating server (i.e., a parameter server). Subsequently, the server aggregates the contributions from the clients, derives an updated model, and distributes the model weights to the participating clients. This process is repeated over several iterations until the global model reaches a certain accuracy level.

Although the raw training data never leave the clients, individual weights may still leak information about the raw data through various means, e.g., privacy inference attacks~\cite{zhaoarxiv2020, SP2019inference, NEURIPS2020inference, SP22inference, CCS21membershipinference}, model inversion attacks~\cite{CCS15inversion, CVPR20modelinversion}, data reconstruction attacks~\cite{USENIX20reconstruction, SP22reconstruction}. These attacks highlight the urgent need for a mechanism capable of securely aggregating the weights computed by the clients for privacy-preserving federated learning (PPFL). An efficient strategy is to introduce multiple non-colluding servers for secure and private aggregation, e.g., Prio~\cite{PrioNSDI2017}, Prio+~\cite{Prio+SCN2022}, ELSA\cite{ELSASP2023}. However, the high communication overhead and requirement of two or more non-colluding servers may not be applicable in practice~\cite{ROFLSP23, EIFFELCCS2022}.

A more promising strategy is the private FL with the single-server setting. Bonawitz et al.~\cite{BonawitzCCS17} introduced the first practical secure aggregation construction for FL, ensuring both security and privacy preservation. This approach further demonstrates high scalability, e.g., supporting thousands of clients, and robustness against client dropouts. In order to mask the private gradient produced during model training, each client generates two types of vectors (self mask and pairwise masks) with the same dimension as the gradients. The self mask is derived from a randomly selected seed for each iteration, while the pairwise masks are derived from temporary session keys negotiated between the client and its neighbors. The summation of the private gradient, self mask and pairwise masks is transmitted to the server. Meanwhile, the secret shares of the seeds are sent to the client's neighbors. The server then asks for the seeds of the survivors and dropouts for mask reconstruction and uses the reconstructed mask to unmask the updated gradient. Therefore, the server learns the summation of the input vectors of many clients securely without obtaining any information beyond the sum. Subsequently, many works inherited and magnified this double-masking approach with improved communication efficiency~\cite{BellCCS20, SP2023flamingo} or enhanced security functionalities~\cite{TIFS2020VerifyNet, SEC2023ACORN}.

However, a significant drawback of these single-server secure aggregation schemes lies in the high round and communication complexities. Since the learning process involves thousands of geographically distributed clients simultaneously~\cite{LOKISP2024}, the costly interactions will cause tremendous communication latency, impeding the efficiency of model training. 
To this end, Ma et al.~\cite{SP2023flamingo} proposed Flamingo, a scheme that substantially reduces communication rounds between the server and clients for each aggregation. In this scheme, the secrets established with key agreement among clients are used to generate pairwise masks for each iteration. Meanwhile, a set of decryptors (also named committee) are selected to decrypt pairwise masks of the clients with the threshold decryption technique. Therefore, the secrets can be reused throughout the collection procedure, significantly reducing the computational and communication costs of key exchange and secret sharing. Nevertheless, Flamingo requires the selection and sharing of self masks in each iteration for privacy preservation, inevitably incurring heavy computational and communication costs.

In addition to efficiency, it is also important to consider the common issue that clients may drop out during interactions due to limited compute and battery power or unstable network conditions. Existing secure aggregation schemes based on the double-masking approach~\cite{BonawitzCCS17, BellCCS20, SP2023flamingo} adopt consistency checking to address this challenge. This check assists clients to reach consensus on individuals who have dropped out, then a malicious server cannot recover both self and pairwise masks for the targeted clients by disseminating counterfeit dropout sets to different clients~\cite{AAAI23so}. However, the necessity of multiple interactions between the server and clients for consistency checks causes considerable computational and communication costs.

Driven by the above limitations, our objective is to \textit{formulate an efficient single-server secure aggregation scheme with fewer communication rounds and smaller overheads without compromising privacy preservation and dropout-robustness}.

\subsection{Our Contributions}
To fulfill our objective, we propose Fluent, a round-efficient single-server secure aggregation scheme for private federated learning. Fluent supports the malicious server setting, and we demonstrate the security of Fluent with a standard simulation-based approach. Our main contributions are as follows:

\noindent\textbf{One-time handshake and secret sharing.} The first key contribution is that we enable clients to mask their private vectors and decryptors to provide secret reconstruction materials without interaction, thereby optimizing communication efficiency. Intuitively, we execute a one-time pre-round phase at the beginning of the model training. Clients perform handshakes for key exchange and private seeds sharing with decryptors only once. Decryptors may repeatedly employ the secret shares of both self masks and pairwise masks across several iterations of aggregation. All these operations will not compromise clients' privacy when the number of clients is below a threshold.

\noindent\textbf{One round consistency check and unmasking.} The second key contribution is that we formulate a new algorithm to accomplish both the consistency check and gradient unmasking in one logical step. Existing schemes~\cite{BellCCS20, SP2023flamingo, ASIACRYPT2023LERNA, MicroFedML} reconstruct seeds and unmask vectors after each consistency check, which results in high latency. In our design, the decryptors furnish the encrypted shares instead of offering reconstruction materials after each consistency check. The server can decrypt the ciphertext, reconstruct seeds, and unmask the private vectors correctly only when the consistency check is achieved. Although this approach can result in limited additional computational costs on the server side, it reduces one round of communication compared to the state-of-the-art schemes. This is a crucial optimization in our distributed setting. We also propose an alternative approach for efficient consistency check. It maintains the same number of communication rounds as existing schemes but decreases the computational and communication costs of one step from $O({n_I}^2)$ to $O(1)$, where ${n_I}$ is the number of decryptors (\S~\ref{section: cross-checking}).

\noindent\textbf{Dynamically joining clients.} Flamingo~\cite{SP2023flamingo} introduces an approach for participants to locally generate their neighbors and establish a common graph with a random string. However, this can only support a constant number of clients ($\mathcal{N}$) participating in a training session, but fails to handle the dropout situation. We propose a method to accommodate dynamic clients and decryptors joining, enhancing the scalability and flexibility of Fluent. This extension includes two modifications: a new client selection algorithm to facilitate dynamic client joining and a multilevel (hierarchical) threshold secret sharing~\cite{MTSS1998, HTSS2007JoC, HarnM14MTSS} to replace the Shamir's secret sharing scheme. In this way, newly added decryptors will receive shares of the long-term secret seeds from clients without leaking the private information. Compared to the intuition of committee evolution~\cite{TCC2020serverless, Crypto2021YOSO} in Flamingo~\cite{SP2023flamingo}, only new clients join the committee, as clients only need to share their seeds with newly added decryptors and existing decryptors do not need to be involved in the process. This will save numerous computational and bandwidth resources.
 
\noindent\textbf{Implementation.} We implement and comprehensively evaluate Fluent from different perspectives. Compared to existing state-of-the-art schemes~\cite{BellCCS20, SP2023flamingo}, Fluent improves the computational cost by at least 75\% and communication overhead by at least 25\% for clients that are not decryptors. Fluent also reduces the communication overhead for the server while only slightly increasing the computational cost. The experimental results also demonstrate that Fluent has good adaptability in the face of an increase in client size, committee size, and dropout rate. Fluent is applicable to scenarios where both lightweight and heavyweight devices simultaneously participate in training tasks.

\section{Preliminaries} \label{notation}
\subsection{Notations}
Let $\mathbb{N}=\{1,2,\dots\}$ be the natural numbers, and $[n]$ be the set of integers $\{1,2,\dots,n\}$. For a finite field $\mathbb{F}$, we use $x\xleftarrow{\$} \mathbb{F}$ to denote uniformly sampling $x$ from $\mathbb{F}$, and $\mathbb{Z}_q$ to denote the finite field with $q$ elements. Vectors are denoted in boldface, e.g., $\mathbf{x} \in \mathcal{X}^l$. $\lambda$ is the security parameter. Our adversaries are non-uniform Probabilistic Polynomial-Time (PPT) ensembles $\mathcal{A} = \{\mathcal{A}_{\lambda}\}_{\lambda \in \mathbb{N}}$. We use $\overset{c}{\approx}$ to denote the computational indistinguishability between two ensembles of distributions: $\{D_\lambda\}_{\lambda \in \mathbb{N}}$ and $\{D'_\lambda\}_{\lambda \in \mathbb{N}}$.

\subsection{System Setting and Threat Model} \label{threat model}
There are two types of entities in our FL setting:
\begin{itemize}[leftmargin=*]
    \item Clients. With an initially untrained global model from the server, clients train the model locally with their respective datasets and obtain a high-dimensional gradient update $\mathbf{x_i}$. A subset of clients $S_t$ is randomly selected for the $t$-th iteration to contribute to model training. Each client communicates with the server through a private and authenticated channel. Messages destined for other clients are forwarded via the server.
    Besides, a designated set of decryptors $\mathcal{I}$ is selected. They store the long-term secret shares of clients. Once receiving a prompt from the server, decryptors reply with secret shares for seed reconstruction and unmasking.
    \item Server. Once collecting enough masked private vectors from clients, the server averages the gradient updates and computes a global update $\mathbf{z}=\sum_{i\in S_t}\mathbf{x_i}$ with the assistance of decryptors. Subsequently, the server updates the model based on the aggregated update $\mathbf{z}$.
\end{itemize}

\noindent\textbf{Threat model.} From a practical standpoint, we assume that the server is malicious but obligated to execute the public key commitment honestly, resembling the semi-malicious setting in~\cite{BellCCS20} and malicious setting in~\cite{SEC2023ACORN}. We also assume a fraction of clients are compromised. They may collude with the malicious server to infer the private vectors and datasets of benign clients. The objective of the proposed scheme is to enable the server to compute the global model while maintaining robustness to a certain fraction of client dropouts and compromises. We impose an upper bound on the fraction of clients and decryptors that may drop out in any given aggregation iteration ($\eta_\mathcal{D}$) and on the fraction of clients that are compromised ($\eta_\mathcal{C}$). For simplicity, we assume that clients and decryptors share an equal probability of dropping out and being compromised. Meanwhile, let $\eta=\eta_\mathcal{D}+\eta_\mathcal{C}$ denote the upper limit for the fraction of a client that drops out or is corrupted.

\subsection{Cryptographic Building Blocks}

\begin{definition}\label{def1}
    \textbf{\textit{(Computational Diffie-Hellman (CDH) Problem).}}
    \textit{Let $\mathbb{G}$ be a cyclic multiplicative group of prime order $q$, $g$ is the generator of $\mathbb{G}$. Given $(g, g^a, g^b)\in \mathbb{G}^3$, where $a,b\xleftarrow{\$}\mathbb{Z}_q^*$, the CDH problem is to compute $g^{ab}\in \mathbb{G}$. The CDH assumption means that the CDH problem is computationally intractable to solve in probabilistic polynomial time.}
\end{definition}

\noindent\textbf{Diffie-Hellman key exchange.} Let $\mathbb{G}$ be a group of order $q$, and $g$ be a generator of $\mathbb{G}$. Alice and Bob can safely establish a shared secret (assuming a passive adversary) as follows.

Alice samples a secret $a \xleftarrow{\$} \mathbb{Z}_q$ and sets her public value to $g^a\in \mathbb{G}$. Bob samples his secret $b \xleftarrow{\$} \mathbb{Z}_q$ and sets his public value to $g^b\in \mathbb{G}$. Alice and Bob exchange public values and raise the value of the other party to their secret, i.e., $g^{ab}=(g^a)^b=(g^b)^a$. If the CDH assumption holds, the shared secret $g^{ab}$ is only known to Alice and Bob.

\noindent\textbf{Pseudorandom generators.} A pseudorandom generator (PRG) is a deterministic function that takes a random seed in $\{0,1\}^{\lambda}$ and produces a longer string. The security of a secure PRG guarantees that its output on a uniformly random seed is computationally indistinguishable from a uniformly sampled element of the output space, as long as the seed is hidden from the distinguisher. For simplicity, we assume that all inputs that exceed the input range will be mapped to $\{0,1\}^{\lambda}$ in advance.

\noindent\textbf{Shamir's secret sharing scheme.} This is an $\kappa$-out-of-$n$ secret sharing scheme $\Pi_{\kappa,n}$ with information-theoretic security, operating over a finite field $\mathbb{F}_q$ for a large prime $q$. Such a scheme consists of two algorithms:
\begin{itemize}[leftmargin=*]
    \item SS.share$(s,\kappa,\mathcal{U}) \rightarrow \{(u,s^{(u)})\}_{u \in \mathcal{U}}$: the sharing algorithm takes as input a secret $s$, a set $\mathcal{U}$ of $n$ field elements, and a threshold $\kappa \leq n$, and outputs a set of shares $\{{(u,s^{(u)})}\}_{u \in \mathcal{U}}$, where each share is associated with a user $u\in \mathcal{U}$.
    \item SS.reconstruct$(\{u,s^{(u)}\}_{u\in \mathcal{I}}) \rightarrow s\,/\perp$: the reconstruction algorithm takes as input the shares and outputs the secret $s$ if $\lvert \mathcal{I} \rvert \geq \kappa$ and each share is a valid share of $s$, otherwise outputs $\perp$.
\end{itemize}

For correctness, it requires that $\forall s\in \mathbb{F}_p$, $\forall \kappa, n$ with $1\leq \kappa \leq n$, $\forall \mathcal{U} \subseteq \mathbb{F}_q$, where $\lvert \mathcal{U} \rvert=n$, if $\{(u$, $s^{(u)})\}_{u \in \mathcal{U}}$ $\leftarrow \mathrm{SS.share}(s,\kappa,\mathcal{U})$, $\mathcal{I} \subseteq \mathcal{U}$ and $\lvert \mathcal{I} \rvert \geq \kappa$, then $\mathrm{SS.reconstruct}(\{u,s^{(u)}\}_{u\in \mathcal{I}})=s$.
For security, it requires that $\forall s_0$, $s_1 \in \mathbb{F}_q$ and any $\mathcal{I}\subseteq \mathcal{U}$ and $\lvert \mathcal{I} \rvert < \kappa$, $\{\{(u,s_0^{(u)})\}_{u \in \mathcal{U}} \leftarrow \mathrm{SS.share}(s_0,\kappa,\mathcal{U}): \{(u,s_0^{(u)})\}_{u \in \mathcal{I}}\} \overset{c}{\approx} \{\{(u,s_1^{(u)})\}_{u \in \mathcal{U}} \leftarrow \mathrm{SS.share}(s_1,\kappa,\mathcal{U}): \{(u,s_1^{(u)})\}_{u \in \mathcal{I}}\}$.

\noindent\textbf{Authenticated encryption.} An authenticated encryption (AE) scheme provides confidentiality and integrity for end-to-end communications. It consists of a triple of algorithms:
\begin{itemize}[leftmargin=*]
    \item AE.KeyGen($\lambda$)$\rightarrow K$: the key generation algorithm takes a security parameter $\lambda$ as input and outputs a secret key $K$.
    \item AE.Enc($K,m$)$\rightarrow C$: the encryption algorithm takes as input a key $K$ and a message $m$, and outputs a ciphertext $C$.
    \item AE.Dec($K, C$)$\rightarrow m\,/\perp$: the decryption algorithm takes as input a key $K$ and a ciphertext $C$, and outputs the original plaintext $m$, or $\perp$ if decryption fails.
\end{itemize}

For correctness, it requires that $\forall K \in \{0,1\}^{\lambda}$ and $\forall m$, $\mathrm{AE.Dec}(K, \mathrm{AE.Enc}(K, m)) = m$. For security, it requires indistinguishability under a chosen plaintext attack (IND-CPA) and ciphertext integrity (IND-CTXT). Informally, a PPT adversary is computationally infeasible to distinguish between fresh encryptions of two different messages or create new valid ciphertexts with a better than negligible advantage.

\noindent\textbf{Digital signature.}
A digital signature (DS) scheme provides message integrity and authentication for end-to-end communications. It consists of the following three algorithms:
\begin{itemize}[leftmargin=*]
    \item DS.KeyGen$(\lambda)\rightarrow (sk,pk)$: The key generation algorithm takes a security parameter $\lambda$ as input, and outputs a secret key $sk$ and a public key $pk$.
    \item DS.Sign$(sk,m)\rightarrow \sigma$: The signing algorithm takes a secret key $sk$ and a message $m$ as input, and outputs a signature $\sigma$.
    \item DS.Verify$(pk,\sigma,m)\rightarrow b$: The signature verification algorithm takes as input a public key $pk$, a signature $\sigma$, and a message $m$, and outputs a bit indicating the legitimacy of the signature.
\end{itemize}

For correctness, it requires that $\forall m$, for all $(sk, pk)\leftarrow$DS.KeyGen($\lambda$), DS.Verify($pk,$ DS.Sign$(sk, m), m$) = 1 is always true. For security, existential unforgeability under a chosen message attack (EUF-CMA security) of a digital signature requires that no PPT adversary, given a fresh public key, can produce a legitimate signature on a message without the secret key.


\section{Round-efficient Secure Aggregation for FL} \label{scheme}

We first describe the high-level idea of Fluent (\S\ref{intuition}), followed by a comprehensive exposition of the detailed protocol (\S \ref{detail scheme}).

\subsection{High-level Idea} \label{intuition}
Fluent improves the performance of FL by minimizing both the communication round and latency for each secure aggregation, by incorporating two main modifications compared with existing schemes~\cite{BellCCS20, SP2023flamingo}.

\noindent\textbf{(1) One-time handshake and secret sharing.} 
In previous secure aggregation schemes based on double-masking~\cite{BonawitzCCS17, BellCCS20, SP2023flamingo}, each client masks their private vector with a self mask and a pairwise mask. The self masks, and in some cases, the pairwise masks, are randomly selected and shared in each iteration to safeguard the privacy of private vectors. We aim to securely reuse the secret shares (of both self masks and pairwise masks) across multiple iterations. This would reduce the computational and communication costs. 

Since the masks are rebuilt for private vector reconstruction in each iteration, it is imperative that the server remains oblivious to information beyond the current iteration. Our first intuition is to utilize function secret sharing (FSS)~\cite{EUROCRYPT2015FSS, CCS2016FSS} and homomorphic secret sharing (HSS)~\cite{CRYPTO2021HSS, CRYPTO2023HSS}. However, the limited participants in FSS and the constrained function classes in HSS cannot satisfy the scalability and security requirements, respectively.

Capitalizing on the linearity of Shamir's secret sharing scheme, we utilize a function with the homogeneity property to aggregate the shares of clients for mask reconstruction. This must be a one-way function to prevent the malicious server from obtaining the concrete values of the shares. Therefore, we can utilize the lightweight components based on the discrete logarithm (DL) assumption and the linearity property of Shamir's secret sharing scheme. While other one-way cryptographic primitives with homogeneity properties may also help, the expensive computational cost may become a bottleneck. Decryptors compute a common generator for each iteration and compute masked shares to hide the secret shares of participants. In this situation, the resource-intensive handshake and secret sharing only need to be implemented once before the recursive gradient collection and update.

According to Pasquini et al.~\cite{PasquiniCCS22eluding}, the malicious server may target a specific client by disseminating inconsistent global models to different clients. Therefore, we utilize the global model received from the server as the parameter for generator computation. This guarantees that an inconsistent model results in incorrect secret reconstruction, so that the privacy of clients is still rigorously preserved.

\noindent\textbf{(2) One-round consistency check and unmasking.} A malicious server may simply give inconsistent views of user dropouts to different clients to recover the private vector $\mathbf{x_i}$ from a target client~\cite{BonawitzCCS17,BellCCS20}. Consequently, it is imperative to establish a consistent view of who has dropped out for decryptors to construct corresponding masks. However, the conventional process consists of two rounds of communications between the server and the clients or decryptors, each requiring a minimum of $O({n_I})$ communications for each round, where ${n_I}$ denotes the number of decryptors. The key insight is that the decryptors only transmit the shares of masks after the consistency check. Can we amalgamate the consistency check process into the share reconstruction process?

To address this question, we leverage the concept of $\kappa$-heavy-hitters~\cite{BonehSP21, CCS2022STAR} to achieve threshold aggregation. Alex et al.~\cite{CCS2022STAR} introduced STAR, a lightweight private threshold aggregation system designed to collect measurements from clients with identical measurement data. Specifically, private data is encrypted using a session key generated from the measurement. The key can only be reconstructed by utilizing the secret sharing scheme when the number of clients submitting effective key shares reaches the threshold. This scheme seamlessly aligns with our objectives. However, the system lacks effective identity authentication and remains vulnerable to Sybil attacks, i.e., any client with the measurement can counterfeit multiple key materials to reduce the complexity of key reconstruction.

Therefore, we address this limitation by integrating the technology of threshold decryption. Each decryptor is assigned a share of the system secret key for resistance to Sybil attacks. Specifically, once receiving a set $\mathcal{U}_\mathcal{S}$ containing multiple signatures from the server, decryptors encrypt the shares of masks with a temporary key and transmit the ciphertext to the server. Meanwhile, each decryptor generates decryption shares for the temporary keys of other decryptors. Only when the decryption shares reach the threshold can the server calculate the temporary key, decrypt the ciphertext, and reconstruct secret seeds for gradient unmasking. In this way, decryptors can transmit all messages to the server at once and then go offline, which will increase practicality substantially.

Collectively, the aforementioned concepts eliminate the requirement for handshakes and secret sharing in every iteration, providing a round-efficient method to improve the practicality of long-term FL. Figure~\ref{fig: main algorithms} illustrates the overall protocol, and we provide a detailed description of each component below.

\subsection{Detailed Protocol} \label{detail scheme}
\begin{figure}[!t]
  \centering{
  \subfigtopskip=2pt
  \subfigbottomskip=2pt 
  \subfigcapskip=-5pt 
  \subfigure[Workflow of Bell et al.'s scheme~\cite{BellCCS20}]{
    \label{fig: workflow1}
    \includegraphics[width=1.4in]{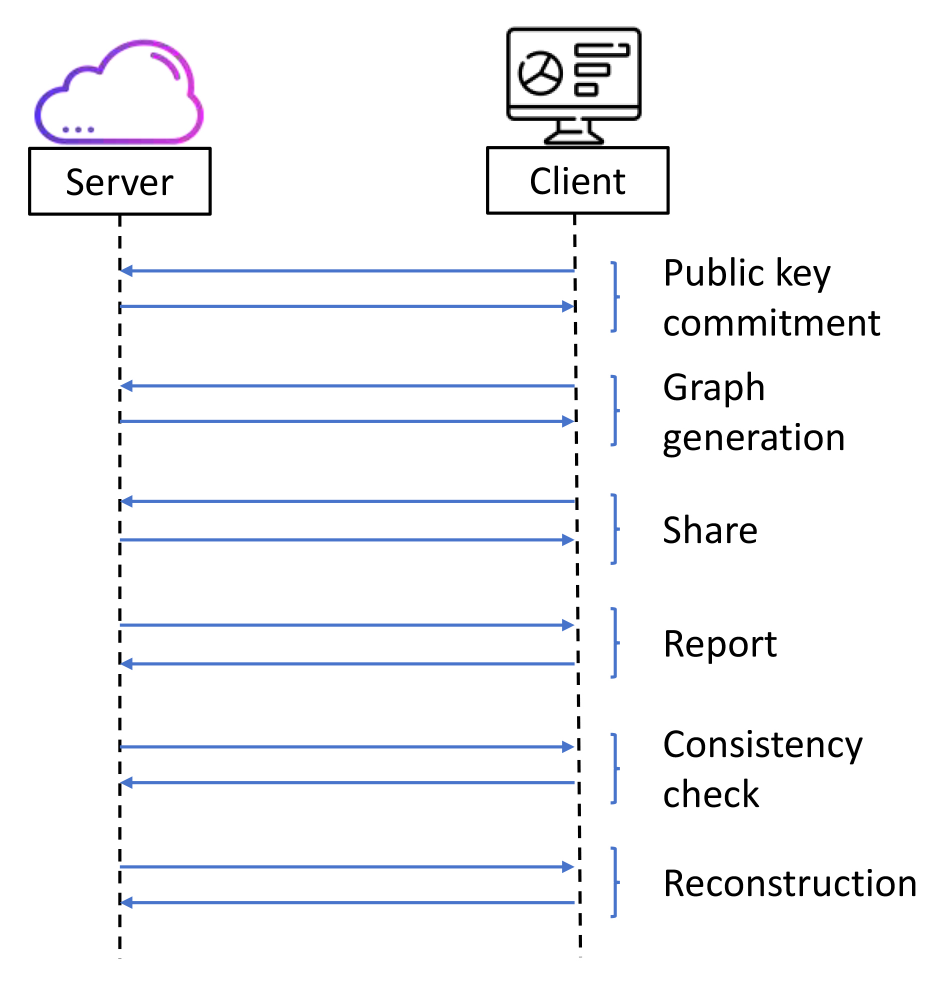}}
  \subfigure[Workflow of Flamingo~\cite{SP2023flamingo} and Fluent]{
    \label{fig: workflow2}
    \includegraphics[width=1.8in]{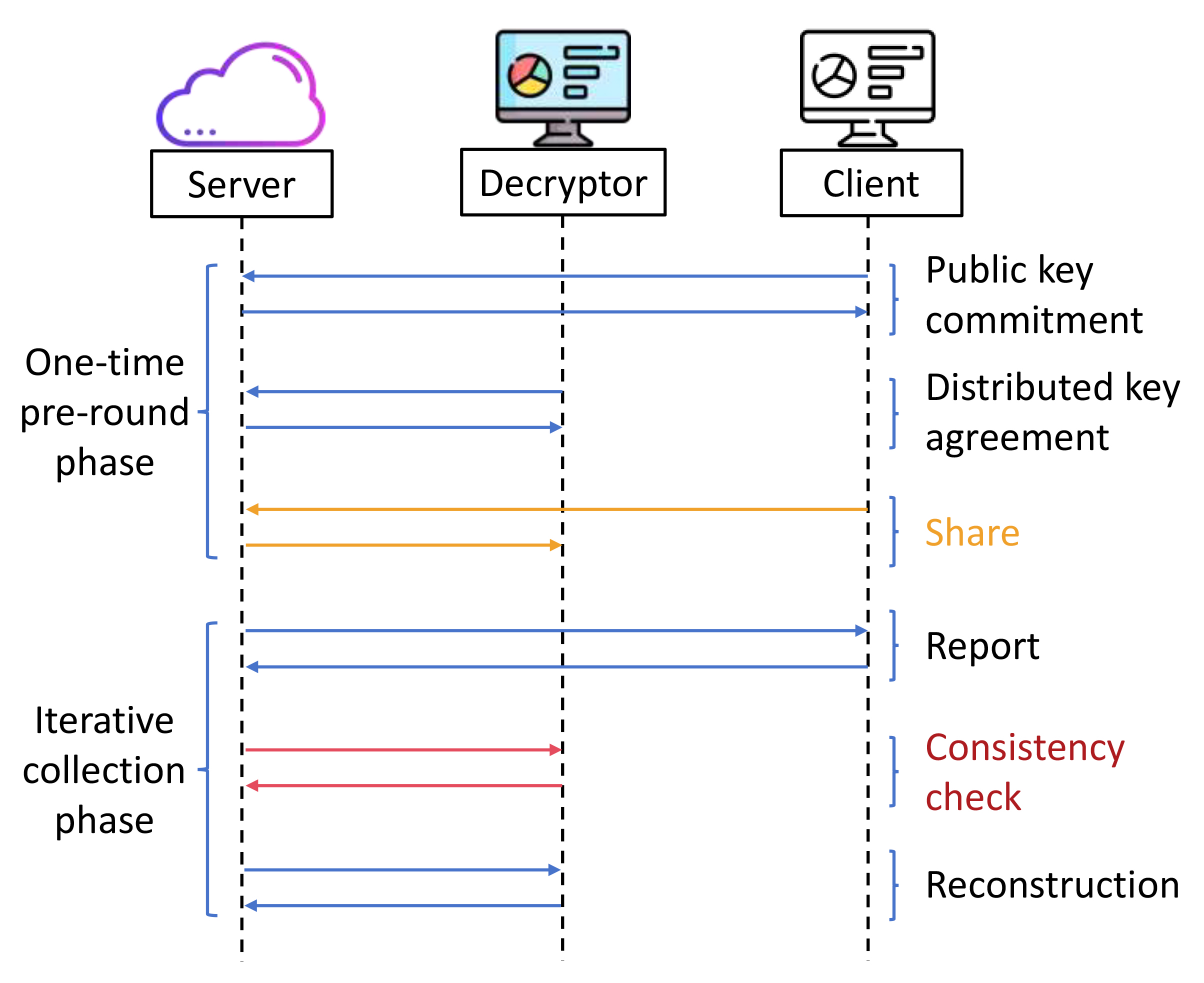}}
  \caption{The workflow of Fluent and existing schemes~\cite{BellCCS20, SP2023flamingo}, where the \textcolor{mycolor}{\textit{share}} step in Figure~\ref{fig: workflow2} is specific to Fluent and \textcolor{red}{\textit{consistency check}} is specific to Flamingo and is not required in Fluent anymore.}
  \label{fig: workflow}
  }
\end{figure}

Our protocol consists of two main phases, i.e., pre-round phase and collection phase. The pre-round phase can be executed only once for all participants. Afterwards, the participants invoke the collection phase for multiple iterations to train a global model.
The workflow of Fluent is presented in Figure~\ref{fig: workflow}. In comparison, Bell et al.'s scheme~\cite{BellCCS20} consists of six iterative rounds, i.e., public key commitments, distributed graph generation, share (mask generation and secret sharing), report, consistency check, and reconstruction. Flamingo~\cite{SP2023flamingo} requires two rounds in the one-time pre-round phase and three rounds in the iterative collection phase (including consistency check). Fluent requires three rounds in the one-time pre-round phase (including share). We defer the full protocol to Figure~\ref{fig: main algorithms} in Appendix.

\subsubsection{Pre-round Phase}
The pre-round phase consists of three one-time components: (1) public key commitments; (2) distributed key generation; and (3) key agreement and seed sharing. The participant and neighbor selection algorithm may adhere to the methodology of Flamingo~\cite{SP2023flamingo}, which allows the server and all clients to compute common targets with a common parameter. Please refer to~\cite{SP2023flamingo} for more detail.

\noindent\textbf{Public key commitments.} A cryptographic accumulator~\cite{CCS22keyaccumulator} is a primitive that produces proofs of membership or non-membership together with verifiable commitments. For simplicity, we utilize the simplest accumulator (i.e., a Merkle tree) to commit the public keys of clients, but several other accumulators are also optional. Alternatively, a trusted public key infrastructure (PKI) is also helpful for credit endorsement. However, the usage of the cryptographic accumulator is compatible with registration-based encryption (RBE)~\cite{RBETCC2018,RBEccs23}, a new cryptographic framework recently introduced as an alternative to identity-based encryption (IBE). RBE may eliminate expensive certificate management and key escrow problems. Meanwhile, the server may take on the responsibility of key curator in RBE, which is a weaker entity than PKI.

For a clear expression, each client $i$ first generates three key pairs $(sk_{i,1},pk_{i,1}),(sk_{i,2},pk_{i,2}),(sk_{i,3},pk_{i,3})$ based on the discrete logarithm and sends $(pk_{i,1},pk_{i,2},pk_{i,3})$ to the server for commitment. The server commits to the public keys and returns the signed root hashes to the client. The specific functions of each key pair are shown as follows:
\begin{itemize}[leftmargin=*]
    \item $(sk_{i,1},pk_{i,1}=g^{sk_{i,1}})$ is used for the generation of pairwise masks between any two clients.
    \item $(sk_{i,2},pk_{i,2}=g^{sk_{i,2}})$ is used for asymmetric cryptographic algorithms, i.e., using $sk_{i,2}$ for signature generation and ciphertext decryption and using $pk_{i,2}$ for signature verification and message encryption.
    \item $(sk_{i,3},pk_{i,3}=g^{sk_{i,3}})$ is only owned by decryptors and used to generate decryption shares of temporary keys for consistency check and unmasking.
\end{itemize}

\noindent\textbf{Distributed key generation.} DKG~\cite{eurocrypt99DKG, iacr2021DKG} enables decryptors to cooperatively generate a master public key $mpk$ without leaking the value of the corresponding master secret key $msk$. After determining the committee of decryptors, they may use DKG for further computation. At the end of DKG, each decryptor $i$ holds a share $msk_i$ of the master secret key $msk$, and the master public key $mpk=g^{msk}$ is publicly known.

\noindent\textbf{Key agreement and seed sharing.} Every two clients first negotiate a pairwise seed with the DH key agreement protocol as $seed_{i,j}=\mathrm{KA.Agree}(sk_{i,1},pk_{i,1})$. Meanwhile, each client also selects a random self mask $seed_i \xleftarrow{\$}\mathbb{Z}_q$. Then, they share both the self mask and pairwise mask with all decryptors $u \in \mathcal{I}$. Specifically, client $i$ computes two sets of shares $(u,seed_i^{(u)})_{u\in \mathcal{I}}\leftarrow \mathrm{SS.Share}(seed_i,\kappa,\mathcal{I})$, $(u,$ $seed_{i,j}^{(u)})_{u\in \mathcal{I}}\leftarrow \mathrm{SS.Share}(seed_{i,j},\kappa,\mathcal{I})$. Then, it generates a session key $key_{i,u}= \mathrm{KA.Agree}(sk_{i,2},pk_{u,2})$ with $sk_{i,2}$ and $pk_{u,2}$. With the session key and an AE scheme, it obtains the corresponding ciphertexts $c_{i,u}=\mathrm{AE.Enc}(key_{i,u},(i\parallel u \parallel seed_i^{(u)} \parallel seed_{i,j}^{(u)}))$ for $j\in A_{i,t}$. The ciphertext, together with an index $(u,c_{i,u})$, will be sent to the server, packed, and forwarded to the decryptor $u$. Then $u$ decrypts the ciphertext with $k_{u,i}$ and stores $seed_i^{(u)}$ and $seed_{i,j}^{(u)}$ if the indexes are correct.

\subsubsection{Collection Phase}
After the pre-round phase, the server and all clients execute the collection phase for multiple iterations until the model converges or reaches a pre-defined accuracy level. The collection phase consists of two steps, described as follows:

\noindent\textbf{Report.} In this step, the server and clients first determine the participants and some disposable parameters for the current iteration of training. Once accepting a global model from the server, each client will generate a masked gradient and transmit the gradient to the server for aggregation.

Take iteration $t$ as an example. Given a global model $\Theta_t$, the server first computes $S_t\leftarrow \mathrm{ChooseSet}(\Theta_t, t, n_t,N)$ as participants for model training and transmits the global model $\Theta_t$ to client $i\in S_t$. Then, client $i$ locally chooses its neighbors $A_{i,t} \leftarrow \mathrm{FindNeighbors}(\Theta_t, S_t,i)$. Both the ChooseSet and FindNeighbors algorithms may follow~\cite{SP2023flamingo}. In particular, both algorithms generate random strings with public strings and parameters and determine the results through the first several bits of the strings. However, we compute the participants and generate the graph with a randomness generated from the global model $\Theta_t$ in that iteration.

Each client $i\in S_t$ trains the model with its local dataset $D_i$ and the global model $\Theta_t$ received from the server and obtains a private vector $\mathbf{x_i}$. Take the model $\Theta_t$ and the iteration $t$ as input, client $i$ and other clients independently compute a common generator as $g_{t}=H_{mtp}(\Theta_t\parallel t)$, where $H_{mtp}$ is a map-to-point hash that maps inputs of arbitrary length to an element of $\mathbb{G}$. Afterwards, client $i$ computes self mask as $\mathbf{r_i}=PRG(g_t^{seed_i})$ and pairwise mask as $\mathbf{m_{i,j}}=PRG(g_t^{seed_{i,j}})$ for $j\in A_{i,t}$. Then, the masked input will be $\mathbf{y_i}=\mathbf{x_i}+\mathbf{sk_i}$, where $\mathbf{sk_i}=\mathbf{r_i}-\sum_{0<j<i} \mathbf{m_{i,j}}+\sum_{i<j\leq n_t}\mathbf{m_{i,j}}$. Finally, client $i$ computes signature $\sigma_{i}\leftarrow \mathrm{DS.Sign}(sk_{i,2},m_{i}=``online'' \parallel i \parallel t)$ for consistency check. The message $\{\mathbf{y_i}, (m_i,\sigma_i)\}$ will be sent to the server.

Due to the same reason explained in the last subsection, we utilize the global model $\Theta_t$ and iteration $t$ to determine the participants $S_t$ and neighbors $g_t$. This can withstand the attacks caused by disseminating inconsistent models~\cite{PasquiniCCS22eluding} and enhance the security of the model training. If inconsistent models are distributed to clients, the participants, generator $g_t$, and seeds will not be reconstructed correctly, and neither will the private vectors. Besides, the distribution of the global model $\Theta_t$ will not result in extra rounds or communication.

\noindent\textbf{Consistency check and unmasking.} The server collects messages for a determined time period. If it receives fewer than $(1-\eta)n_t$, it aborts. Otherwise, it defines a global set of dropouts $\mathcal{U}_\mathcal{D}$ and a set of survivors $\mathcal{U}_\mathcal{S}$. It then sends the message and signature pairs $(m_i,\sigma_i)$ to all decryptors together with the sets $\mathcal{U}_\mathcal{S}$ and $\mathcal{U}_\mathcal{D}$.

Each decryptor $u\in \mathcal{I}$ first checks the legitimacy of the sets and signatures. It aborts if any of these checks fail or there are less than $\kappa$ valid signatures. Otherwise, it randomly selects $k_u\xleftarrow{\$}\mathbb{Z}_q$ and computes $c_{seed,u}=$ $\mathrm{AE.Enc}(k_u,$ $ g_{t}^{seed_i^{(u)}} \parallel g_{t}^{seed_{j,k}^{(u)}})_{i\in \mathcal{U}_\mathcal{S},j\in \mathcal{U}_\mathcal{D},k\in A_{j,t}}$, $c_{key,u}=$ $Encode(k_u)\cdot mpk^{sk_{u,3}+H(\mathcal{U}_\mathcal{S}\parallel \mathcal{U}_\mathcal{D})}$, where $Encode$ encodes $k_u$ to an element of $\mathbb{G}$ and can be decoded to $k_u$ with $Decode$. Then, decryptor $u$ further computes decryption shares $c_{u,i}={(g^{H(\mathcal{U}_\mathcal{S} \parallel \mathcal{U}_\mathcal{D})}\cdot pk_{i,3})}^{msk_u}$ for $i\in \mathcal{I} \backslash \{u\}$ for all other decryptors and sends the message $\{c_{seed,u},c_{key,u},c_{u,i}\}_{i\in \mathcal{I} \backslash \{u\}}$ to the server.

Once collecting enough messages from decryptors, the server can decrypt the ciphertext $c_{key,u}$ and obtain the secret key of neighbor $u$ with $k_u=Decode(c_{key,u}/\Pi_{i\in \mathcal{I} \backslash \{u\}}{c_{i,u}}^{\beta_i})$, where $\beta_i$ is the Lagrange coefficient. Then, the secret shares of clients can be calculated as $\{g_{t}^{seed_i^{(u)}}\parallel g_{t}^{seed_{j,k}^{(u)}}\}_{i \in \mathcal{U}_\mathcal{S},j\in \mathcal{U}_\mathcal{D}, k\in A_{j,t}} = \mathrm{AE.Dec}(k_u,c_{seed,u})$.

Thereafter, the server computes the masks as $g_t^{seed_i}=\Pi_{u=1}^{\kappa}({g_{t}^{seed_i^{(u)}}})^{\beta_u}$ and $g_{t}^{seed_{j,k}}=\Pi_{u=1}^{\kappa}({g_{t}^{seed_{j,k}^{(u)}}})^{\beta_u}$, where $u\in \mathcal{I}$. Then, the server can unmask the private vectors and obtain the aggregated model as $\mathbf{z}=\sum_{i\in \mathcal{U}_\mathcal{S}} \mathbf{x_i}=\sum_{i\in \mathcal{U}_\mathcal{S}}\mathbf{y_i}-\mathbf{sk}$, where $\mathbf{sk}=\sum_{i\in \mathcal{U}_\mathcal{S}}PRG(g_t^{seed_i})\pm \sum_{i\in \mathcal{U}_\mathcal{D}}PRG(g_t^{seed_{j,k}})$.

\vspace{\baselineskip}
\noindent\textbf{Correctness.}
\begin{align*}
  k_u &= Decode(c_{key,u}/(\Pi_{i\in \mathcal{I}\backslash \{u\}}{c_{i,u}}^{\beta_i})) \\
    &= Decode(c_{key,u}/(\Pi_{i\in \mathcal{I}\backslash \{u\}}{g}^{(H(\mathcal{U}_\mathcal{S} \parallel \mathcal{U}_\mathcal{D})+sk_{u,3})msk_i \beta_i})) \\
    &= Decode(c_{key,u} / {g}^{(H(\mathcal{U}_\mathcal{S} \parallel \mathcal{U}_\mathcal{D})+sk_{u,3})\sum_{i\in \mathcal{I}\backslash \{u\}}msk_i\beta_i}) \\
    &= Decode(c_{key,u}/ g^{(H(\mathcal{U}_\mathcal{S} \parallel \mathcal{U}_\mathcal{D})+sk_{u,3}) msk}) \\
    &=Decode(c_{key,u}/mpk^{H(\mathcal{U}_\mathcal{S} \parallel \mathcal{U}_\mathcal{D})+sk_{u,3}}) \\
    &=Decode(Encode(k_u)) \\
    &=k_u
\end{align*}

\section{Security Analysis}  \label{security proof}
We first provide a formal definition of the ideal functionality of Fluent, followed by the main theorems regarding the privacy preservation and security of Fluent.

Let $\Pi$ denote the protocol of Figure~\ref{fig: main algorithms}. We denote the functionality of $\Pi$ by $\mathcal{F}_{\mathbf{x},T,\mathcal{N},\eta,\kappa}^{Fluent}$ as it is parameterized by the private vector $\mathbf{x}$ of the set $\mathcal{N}$ with a fraction $\eta$ of compromised or dropout clients within $T$ iterations and a threshold $\kappa$. It takes as input a partition of honest clients (a collection of pairwise disjoint subsets $\{S_1, S_2, \dots, S_T\}$). For each subset $S_t$, it returns $\sum_{i\in S_t}\mathbf{x_i}$ if the subset is ``large enough'', i.e., $\lvert S_t \rvert \geq (1-\eta)\lvert \mathcal{N} \rvert$, or answers $\perp$ otherwise.

\begin{definition}\label{def2}
    \textbf{\textit{(Idea functionality of $\Pi$).}}
Let $q,l,T$ be integers, $\{S_t\}_{t\leq T} \subseteq \mathcal{N}$ be the set of $\mathcal{N}$, and $\eta_\mathcal{C}, \eta_\mathcal{D}$ be the fraction of clients being compromised and dropping out, respectively.
Given a set $S$, a set of clients $S_t$, a set of clients that drop out $\mathcal{U_D}$ such that $\lvert \mathcal{U_D}\rvert/\lvert S_t\rvert \leq \eta_\mathcal{D}$, a set of compromised clients $\mathcal{U_C}$ such that $\lvert \mathcal{U_C}\rvert/\lvert S_t\rvert \leq \eta_\mathcal{C}$, and $\mathcal{X}_{\mathcal{U}}=\{\mathbf{x_i}\}_{i\in \mathcal{U}}$, where $\mathbf{x_i}\in \mathbb{X}^l$, the ideal functionality of $\Pi$ denoted by $\mathcal{F}_{\mathbf{x},T,\mathcal{N},\eta,\kappa}^{Fluent}$, is defined as

\begin{equation*}
\mathcal{F}_{\mathbf{x},T,\mathcal{N},\eta, \kappa}^{Fluent}(\mathcal{S})=\left\{
 \begin{array}{ll}
    \sum_{i\in \mathcal{S}} \mathbf{x_i}  & \makecell{\mathcal{S}\subseteq S_t\backslash \mathcal{U_D} \backslash \mathcal{U_C}~\mathrm{and}\\ \lvert \mathcal{S} \rvert \geq (1-\eta_\mathcal{D})\lvert S_t \rvert},   \\
    \perp  & \mathrm{otherwise.}
\end{array}
\right.
\end{equation*}
\end{definition}

\begin{lemma}\label{lemma1}
Let $G=(V,E)$ be a connected graph and $\{\mathbf{x_i}\in \mathbb{X}_R^l\}_{i\in V}$. For all $l$, the following two distributions are indistinguishable:

\begin{equation*}
 \begin{aligned}
 &\left\{
  \begin{array}{c}
    (\mathbf{x_i}+\sum_{j \in N(i)}\mathbf{r_{i,j}}): \\
    \forall (i,j)\in E, 
    \begin{array}{c}
       \mathbf{r_{i,j}} \leftarrow \mathbb{X}_R^l,~\mathrm{if}~i<j   \\
        \mathbf{r_{i,j}}=-\mathbf{r_{j,i}},~\mathrm{if}~i>j
    \end{array}
\end{array}
\right\}_{i\in V} \\
    &\overset{c}{\approx}
\left \{\mathbf{w_i}\leftarrow \mathbb{X}^l: \sum_{i\in V}\mathbf{w_i}=\sum_{i\in V}\mathbf{x_i}
\right \}_{i\in V}.
\end{aligned}
\end{equation*}
\end{lemma}

\begin{theorem}\label{the1}
    \textbf{(Privacy preservation of $\Pi$).} If the assumptions of DL and CDH hold, given $\{g_t\}_{t\in [T]}\in \mathbb{G}$, and $\{g_t^{sk}, g_t^{sk_u}\}_{t\in [t_1],u\in [\kappa-1]}\in \mathbb{G}$, where $\{(u,sk_u)\}_{u\in \mathcal{N}}\leftarrow \mathrm{SS.share}(sk,\kappa,\mathcal{N})$, $t_1<T$, and $sk\in \mathbb{Z}_q$, it is computationally intractable for a PPT adversary $\mathcal{A}$ defined in \S\ref{threat model} to calculate $sk$, $g_{t'}^{sk}$, or $g_{t'}^{sk_u}$ for any $t_1<t'\leq T,u\in[\kappa-1]$.
\end{theorem}

\noindent\textit{Proof:} We refer readers to Appendix~\ref{proof: theorem 1} for detailed proof.

\begin{theorem}\label{the2}
\textbf{(Security of $\Pi$).} Assume the existence of a EUF-CMA signature scheme, a secure DH key agreement scheme, a secure PRG and a symmetric AE scheme that is IND-CPA and IND-CTXT. There exists a PPT simulator $\rm{Sim}$ such that, for all $\mathcal{U_C}, \mathcal{U_D}\subset S_t$, $\lvert \mathcal{U_C} \rvert \leq \eta_\mathcal{C} \lvert S_t \rvert$, $\lvert \mathcal{U_D} \rvert \leq \eta_\mathcal{D} \lvert S_t \rvert$, input $\mathcal{X}=\{\mathbf{x_i}\}_{i\in S_t\backslash \mathcal{U_C} \backslash \mathcal{U_D}}$, $2\kappa>(1+\eta_\mathcal{C}-\eta_\mathcal{D})n_{I}$, and for any malicious adversary $\mathcal{A}$ controlling the server and the set of corrupted clients $\mathcal{U_C}$ that honestly executes the public key commitment, the output of $\rm{Sim}$ is computationally indistinguishable from the joint view of the server and the corrupted client $\mathrm{Real}(\mathcal{U_C})$, i.e., $\mathrm{Real}(\mathcal{U_C}) \overset{c}{\approx} \mathrm{Sim}_{\mathcal{F}_{\mathbf{x},T,\mathcal{N},\eta,\kappa}^{Fluent}}(\mathcal{U_C})$, where the simulator can query once the ideal functionality $\mathcal{F}_{\mathbf{x},T,\mathcal{N},\eta,\kappa}^{Fluent}$.
\end{theorem}

\noindent\textit{Proof:} We refer readers to Appendix~\ref{proof: theorem 2} for detailed proof.

\renewcommand\arraystretch{1.8}
\begin{table*}[htbp]
\caption{The computational and communication asymptotics of various secure aggregation algorithms. $N$ is the total number of clients, and $n_t$ is the number of clients chosen to participate in iteration $t$. The number of decryptors is ${n_I}$. Let $A_t$ be the upper bound on the number of neighbors of a client, and let $l$ be the dimension of client’s input vector.}
\begin{center}
\resizebox*{\linewidth}{!}{
\begin{tabular}{ccccccc}
\hline
\multirow{2}{*}{Properties} &  \multicolumn{2}{c}{\textbf{Bell et al.~\cite{BellCCS20}}}  & \multicolumn{2}{c}{\textbf{ Flamingo~\cite{SP2023flamingo} }} & \multicolumn{2}{c}{\textbf{  \textbf{Fluent} }}  \\

\cline{2-7}
    &  \textbf{Client} & \textbf{Server}  &  \textbf{Client} & \textbf{Server}  &  \textbf{Client} & \textbf{Server}  \\
\hline
Round &  \multicolumn{2}{c}{6}  & \multicolumn{2}{c}{\makecell{Normal: 1 \\ Decryptor: 2}} & \multicolumn{2}{c}{\makecell{Normal: 1 \\ Decryptor: 1}}  \\

\hline
Computation \\
\cline{1-1}
Pre-round & -- & -- & Decryptor: $\mathcal{O}({n_I}^2)$ & -- & 
\makecell{Normal: $\mathcal{O}(A_t{n_I}^2)$ \\ Decryptor: $\mathcal{O}({n_I}^2)$} & --  \\
Collection  &  $\mathcal{O}(\mathrm{log}^2\,n_t)$  &   
$\mathcal{O}(n_t\mathrm{log}^2\,n_t)$  & \makecell{Normal: $\mathcal{O}(A_t+{n_I}^2)$ \\ Decryptor: $\mathcal{O}({n_I}+n_tA_t)$}   & $\mathcal{O}(n_t^2+A_tn_t{n_I}^2)$    
& \makecell{Normal:$\mathcal{O}(A_t)$  \\ Decryptor: $\mathcal{O}(n_tA_t+{n_I})$ }   &  $\mathcal{O}({n_I}^3+n_tA_t{n_I}^2)$ \\
\hline

Communication  \\
\cline{1-1}
Pre-round & -- & -- &  \makecell{Normal: $\mathcal{O}(1)$ \\ Decryptor:$\mathcal{O}({n_I})$} & $\mathcal{O}(N^2{n_I})$ & 

\makecell{Normal: $\mathcal{O}(N{n_I})$ \\ Decryptor:$\mathcal{O}({n_I})$}  & $\mathcal{O}(N^2{n_I})$  \\
Collection & $\mathcal{O}(\mathrm{log}\,n_t+l)$   &  $\mathcal{O}(n_t(\mathrm{log}\,n_t+l))$ &  \makecell{Normal: $\mathcal{O}(l+A_t+{n_I})$ \\ Decryptor:$\mathcal{O}({n_I}+ n_tA_t)$}  &  $ \mathcal{O}(n_t(l+{n_I}+A_t))$   

& \makecell{Normal: $\mathcal{O}(l)$ \\ Decryptor: $\mathcal{O}(n_tA_t)$}  &  $\mathcal{O}(n_tl+n_tA_t{n_I}+{n_I}^2)$  \\
\hline
\end{tabular}
}
\label{asymptotics comparison}
\end{center}
\end{table*}

\section{Efficiency Analysis}
The numerical boundary conditions of Fluent and state-of-the-art schemes~\cite{BellCCS20, SP2023flamingo} are shown in Table~\ref{asymptotics comparison}.

\noindent\textbf{Communication round.} As depicted in Figure~\ref{fig: workflow}, one of the notable advantages of Fluent over existing PPFL schemes is that it requires the fewest communication rounds for one secure aggregation of model training. The state-of-the-art schemes, i.e., Bell et al.'s scheme~\cite{BellCCS20} and Flamingo~\cite{SP2023flamingo}, need 6 and 3 rounds for one secure aggregation, respectively. In contrast, Fluent achieves the same objective with merely 2 rounds: one is for private vector collection between the server and clients, and the other is for consistency check and unmasking between the server and decryptors. The communication delay, coupled with the need for the server to await a sufficient number of messages for subsequent processing (i.e., mask reconstruction and vector unmasking), would otherwise be increased. Therefore, the decrease in communication rounds is of paramount practical significance due to the widespread and geographically distributed nature of the clients.

\noindent\textbf{Computational cost.}
In Bell et al.'s scheme~\cite{BellCCS20}, each client performs $\leq 5\mathrm{log}\,n_t$ key agreements, encryptions, and signatures generation and verifications ($\mathcal{O}(\mathrm{log}\,n_t)$ complexity), two Shamir's secret shares ($\mathcal{O}(\mathrm{log}^2\,n_t)$ complexity), and one private vector generation ($\mathcal{O}(\mathrm{log}\,n_t)$ complexity). The server executes $n_t$ secret reconstruction key constructions ($\mathcal{O}(n_t\mathrm{log}^2\,n_t)$ complexity) and vector unmasking ($\mathcal{O}(n_t\mathrm{log}\,n_t)$ complexity).

In Flamingo, each decryptor performs DKG ($\mathcal{O}({n_I}^2)$ complexity) in the pre-round phase. In the collection phase, each client generates a masked vector ($\mathcal{O}(A_t)$ complexity), performs one secret shares of self seed ($\mathcal{O}({n_I}^2)$ complexity), $A_t$ encryption of pairwise seeds ($\mathcal{O}(A_t)$ complexity). Decryptors further perform $n_t+{n_I}$ signature verifications for consistency check ($\mathcal{O}(n_t+{n_I})$ complexity), $(1-\eta_\mathcal{D})n_t$ decryptions of self seeds ($\mathcal{O}(n_t)$ complexity), and $\eta_\mathcal{D}n_tA_t$ decryptions of pairwise seeds ($\mathcal{O}(n_tA_t)$ complexity). The server performs $(1-\eta_\mathcal{D})n_t$ self seed reconstruction ($\mathcal{O}(n_t^2)$ complexity), $\eta_\mathcal{D}n_tA_t$ threshold decryptions ($\mathcal{O}(n_tA_t{n_I}^2)$ complexity), and vector unmasking ($\mathcal{O}(n_tA_t)$ complexity).

In Fluent, each client performs $3N$ key agreements ($\mathcal{O}(N)$ complexity), secret sharing to decryptors ($\mathcal{O}(A_t{n_I}^2)$ complexity) in the pre-round phase and each decryptor performs DKG ($\mathcal{O}({n_I}^2)$ complexity) in the pre-round phase. In the collection phase, each client generates one masked vector ($\mathcal{O}(A_t)$ complexity) and one signature ($\mathcal{O}(1)$ complexity) in the collection phase. The decryptor further executes $(1-\eta_\mathcal{D})n_t+\eta_\mathcal{D} n_tA_t$ times symmetric encryption ($\mathcal{O}(n_tA_t)$ complexity), one secret key encryption ($\mathcal{O}(1)$ complexity), and secret key commitment ($\mathcal{O}({n_I})$ complexity). The server executes ${n_I}$ secret key constructions ($\mathcal{O}({n_I}^3)$ complexity), decryption of seed shares ($\mathcal{O}(n_tA_t{n_I})$ complexity), reconstruction of self and pairwise masks ($\mathcal{O}(n_tA_t{n_I}^2)$ complexity), and vector unmasking ($\mathcal{O}(n_tA_t)$ complexity).

\noindent\textbf{Communication overhead.}
In~\cite{BellCCS20}, each client performs $\leq 5\mathrm{log}\,n_t$ key agreements ($\mathcal{O}(\mathrm{log}\,n_t)$ messages), sends $2\mathrm{log}\,n_t$ encrypted shares ($\mathcal{O}(\mathrm{log}\,n_t)$ messages), sends a masked input ($\mathcal{O}(l)$ complexity), and sends $\leq 5\mathrm{log}\,n_t$ signatures, and reveal up to $\leq 2\mathrm{log}\,n_t$ shares ($\mathcal{O}(\mathrm{log}\,n_t)$ message). The server receives or sends $\mathcal{O}(n_t(\mathrm{log}\,n_t+l)$ to each client.

In Flamingo~\cite{SP2023flamingo}, each client sends one message for public key commitment ($\mathcal{O}(1)$ message). Each decryptor sends ${n_I}$ shares for DKG ($\mathcal{O}({n_I})$ messages) in the pre-round setup phase. In the collection phase, each client sends a masked input ($\mathcal{O}(l)$ complexity), and sends ${n_I}$ messages for self seed share, $A_t$ ciphertexts and signatures for pairwise seed decryption ($\mathcal{O}(A_t)$ message) in the collection phase. Each decryptor further sends ${n_I}$ signatures for consistency check ($\mathcal{O}({n_I})$ messages), sends $(1-\eta_\mathcal{D})n_t$ messages ($\mathcal{O}(n_t)$ messages) for self seed reconstruction, and sends $\eta_\mathcal{D} n_tA_t$ messages for pairwise seed reconstruction ($\mathcal{O}(n_tA_t)$ messages) in the collection phase. The server receives or sends $\mathcal{O}(N^2{n_I})$ messages in the pre-round phase and $\mathcal{O}(n_t(l+{n_I}+A_t))$ messages in the collection phase.

In Fluent, each client sends three public keys for public key commitment ($\mathcal{O}(1)$ messages) and sends $N$ ciphertexts to each decryptor ($\mathcal{O}(N{n_I})$ messages) for secret sharing. Each decryptor sends ${n_I}$ messages for DKG  ($\mathcal{O}({n_I})$ messages) in the pre-round setup phase. In the collection phase, each client sends a masked input ($\mathcal{O}(l)$ complexity) and one message and signature pair ($\mathcal{O}(1)$ message) in the collection phase. Each decryptor further sends ciphertext of keys ($\mathcal{O}(1)$ message), ciphertext of seeds ($\mathcal{O}(n_tA_t)$ messages), and commitments of other clients ($\mathcal{O}({n_I})$ messages) in the collection phase. The server receives or sends $\mathcal{O}(N^2{n_I})$ messages in the pre-round phase and $\mathcal{O}(n_tl+n_tA_t{n_I}+{n_I}^2)$ messages in the collection phase.

Theoretically, compared to Bell et al's scheme~\cite{BellCCS20}, both Flamingo~\cite{SP2023flamingo} and Fluent have a high computational complexity on the server side. However, Fluent has the lowest communication complexity and computational complexity. Considering the disparity between the number of clients and decryptors, Fluent is practical and has potential for resource-limited scenarios, e.g., smart home and internet of things (IoT) devices~\cite{IOTJ2022IOT}.


\section{Evaluation} \label{performance}
\subsection{Implementation}
We implement Fluent together with two state-of-the-art schemes~\cite{BellCCS20, SP2023flamingo} to conduct a comparison among them. Some libraries or tools utilized to implement the schemes are shown as follows:
\begin{itemize}[leftmargin=*]
    \item PyCryptodomex 3.17\footnote{https://pypi.org/project/pycryptodomex/}. PyCryptodome is a cryptographic library for Python that provides different cryptographic algorithms and protocols. Algorithms including symmetric encryption algorithm AES in the GCM mode, digital signature algorithm DSA, PRG (AES in counter mode), modular exponentiation, etc. are leveraged in implementation.
    \item Agent-Based Interactive Discrete Event Simulation (ABIDES)\footnote{https://github.com/abides-sim/abides}. ABIDES is an open-source, highfidelity simulator designed for AI research in financial markets. ABIDES is a great fit due to its scalability and ease of configuration.
    \item PyTorch\footnote{https://pytorch.org/}. PyTorch is an open-source machine learning library designed for machine learning and deep learning tasks. The machine learning components and databases are utilized for FL construction and training.
\end{itemize}

We implement all the schemes in the malicious server setting. All servers and clients operate on a PC equipped with a 64-bit Windows 11 OS and a 13th-generation Intel (R) CoreTM i5-13500H processor boasting 16 CPUs, operating at a frequency of 2.6 GHz, and complemented by 32 GB of RAM. Client inputs are 1.6e4-dimensional vectors, each entry being a 32-bit value. All cryptographic primitives are instantiated to achieve the security level of 128 bits.

\subsection{Computational and Communication Cost}
We evaluate the important elements that affect the performance of schemes, i.e., the number of clients, the size of the committee, and the dropout rate.  We refer readers to Table~\ref{table: Computational cost comparison} in Appendix~\ref{appendix: more results} for detailed computational and communication costs of different schemes on both server and client sides, where all experimental results are averaged over 20 rounds of a single aggregation.

\subsubsection{Impact of Number of Clients}
\begin{figure*}[!t]
  \centering{
  \subfigtopskip=2pt
  \subfigbottomskip=2pt 
  \subfigcapskip=-5pt 
  \subfigure[Computational cost of the server]{
    \label{client size sever computation}
    \includegraphics[height=1.25in, width=1.7in]{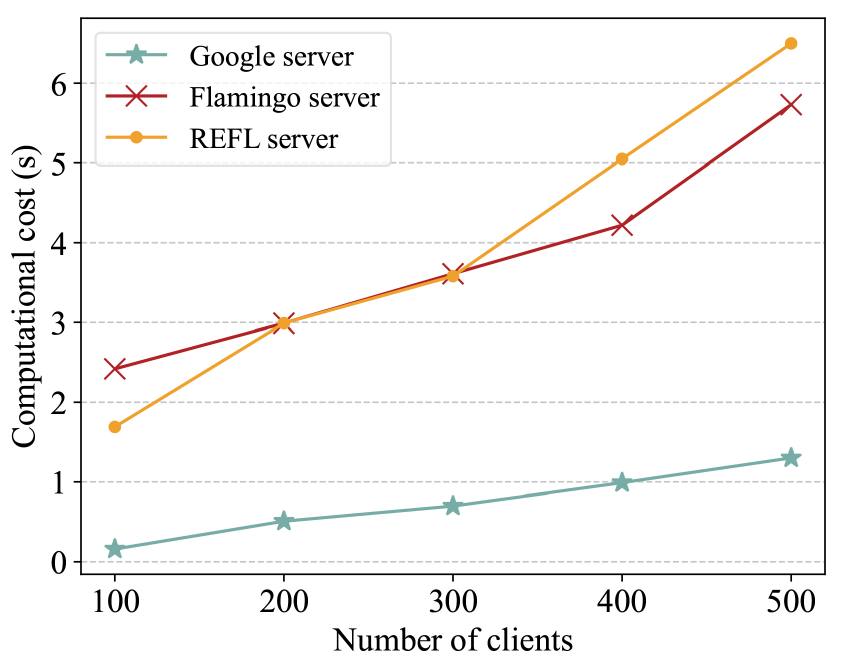}} 
  \hspace{-0.4em}
  \subfigure[Computational cost of clients]{
    \label{client size client computation}
    \includegraphics[height=1.25in, width=1.7in]{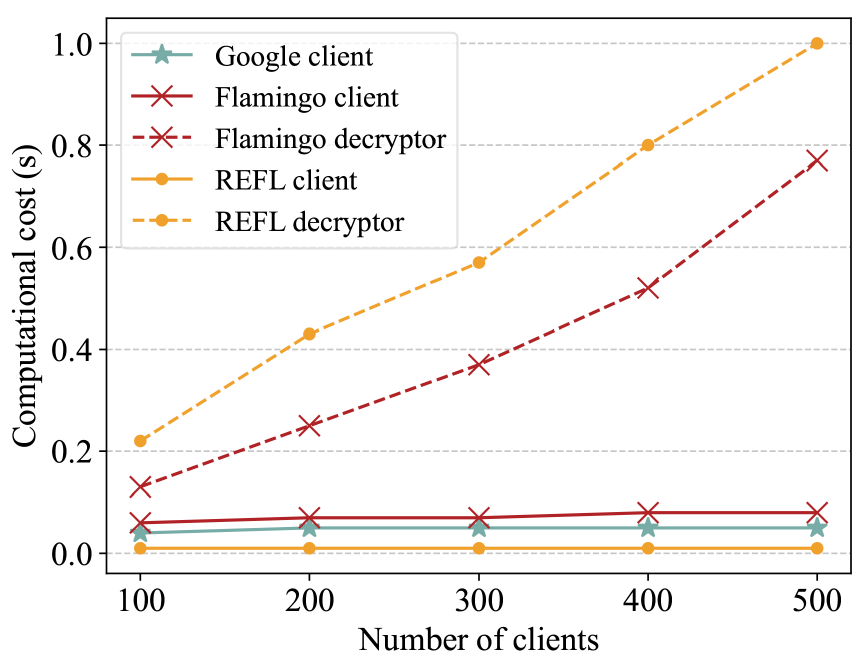}} 
      \hspace{-0.4em}
    \subfigure[Communication overhead of the server]{
    \label{client size sever communication}
    \includegraphics[height=1.25in, width=1.7in]{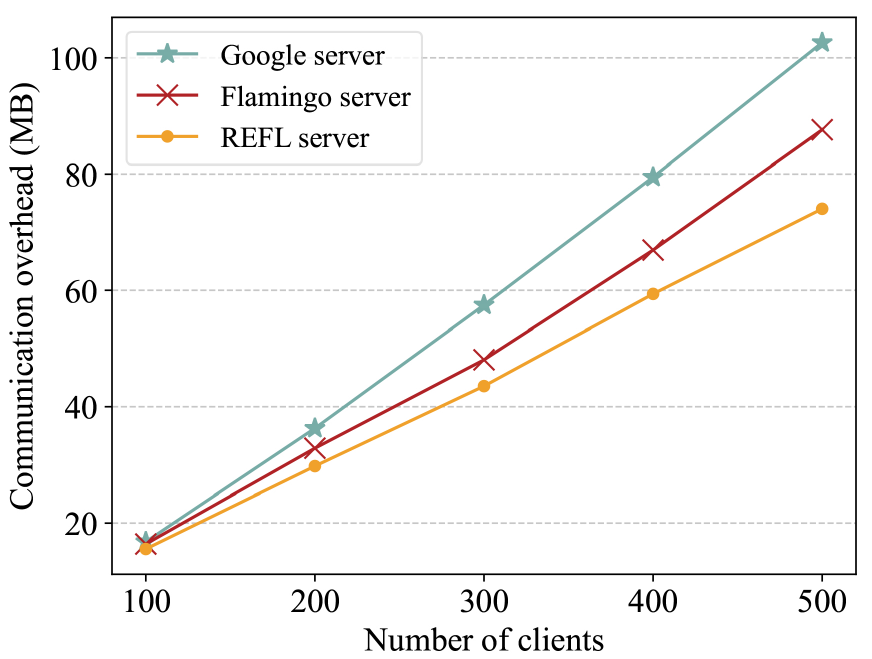}} 
  \hspace{-0.4em}
  \subfigure[Communication overhead of clients]{
    \label{client size client communication}
    \includegraphics[height=1.25in, width=1.7in]{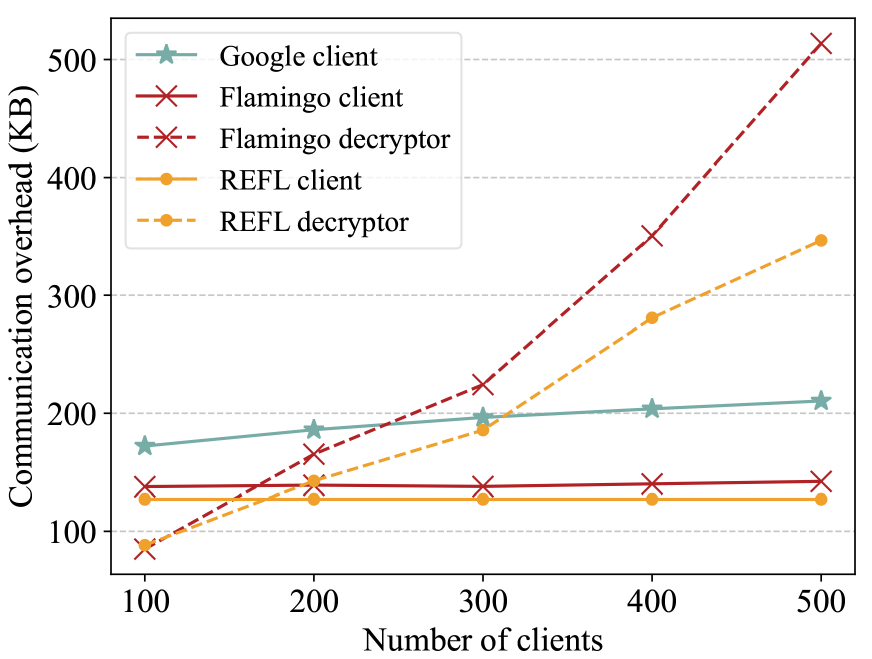}}
  \caption{Computational and communication costs in different schemes in terms of the number of clients.}
  \label{client size figure}
  }
\end{figure*}

The quantity of participants engaged in the secure aggregation is one of the most essential factors for assessing the scalability and flexibility of a FL scheme. We perform different schemes to estimate how the factor affects the schemes. The computational and communication costs across diverse configurations are illustrated in Figure~\ref{client size figure}, where the committee size is fixed at 40 and the dropout rate is fixed at approximately 5\%.

The observed trend reveals that the computational cost of all servers in different schemes generally increases linearly with the growth of the number of clients. On the client side, clients in all three schemes demonstrate consistent stability or only slight growth. This arises from the fact that clients in Fluent always compute a masked vector and a signature that commits to its survivor. The computational cost of clients in Flamingo and Bell et al's scheme correlates solely with the number of decryptors and neighbors, respectively. Conversely, decryptors in both schemes exhibit a linear increase, with a similar magnitude of growth corresponding to the increase in the number of clients. Regarding communication overhead, Fluent experiences a slower increase on both the server and decryptor sides than Flamingo. Relatively, the communication cost of the other two solutions experiences a faster increase as the number of clients escalates. This can be attributed to the design of the pre-round setup, which effectively mitigates extensive redundant communication in each iteration during secure aggregation.

\subsubsection{Impact of Committee Size}
\begin{figure*}[!t]
  \centering{
  \subfigtopskip=2pt
  \subfigbottomskip=2pt 
  \subfigcapskip=-5pt 
  \subfigure[Computational cost of the server]{
    \label{committee size sever computation}
    \includegraphics[height=1.25in, width=1.65in]{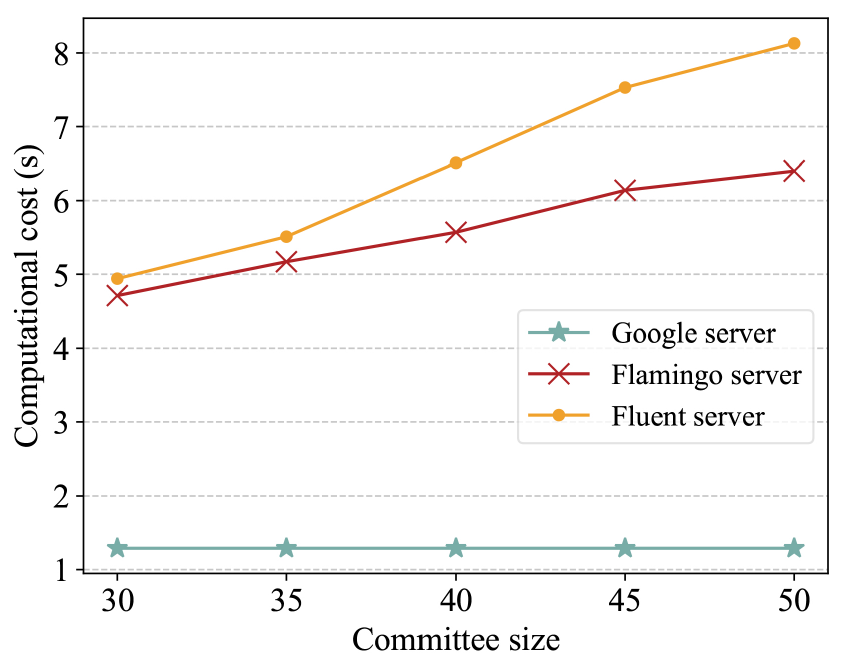}}
  \subfigure[Computational cost of clients]{
    \label{committee size client computation}
    \includegraphics[height=1.25in, width=1.7in]{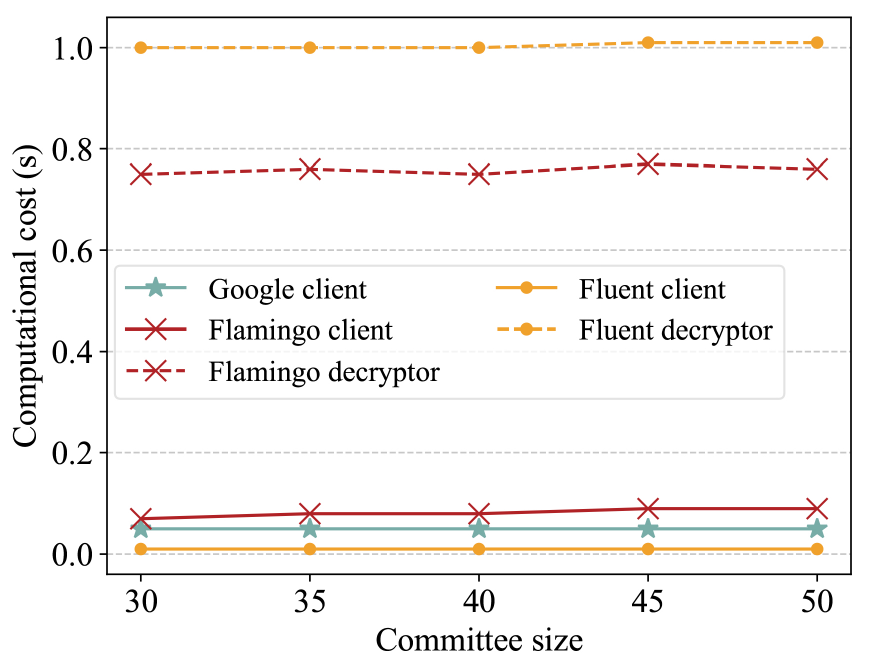}}
      \hspace{-0.4em}
    \subfigure[Communication overhead of the server]{
    \label{committee size sever communication}
    \includegraphics[height=1.25in, width=1.7in]{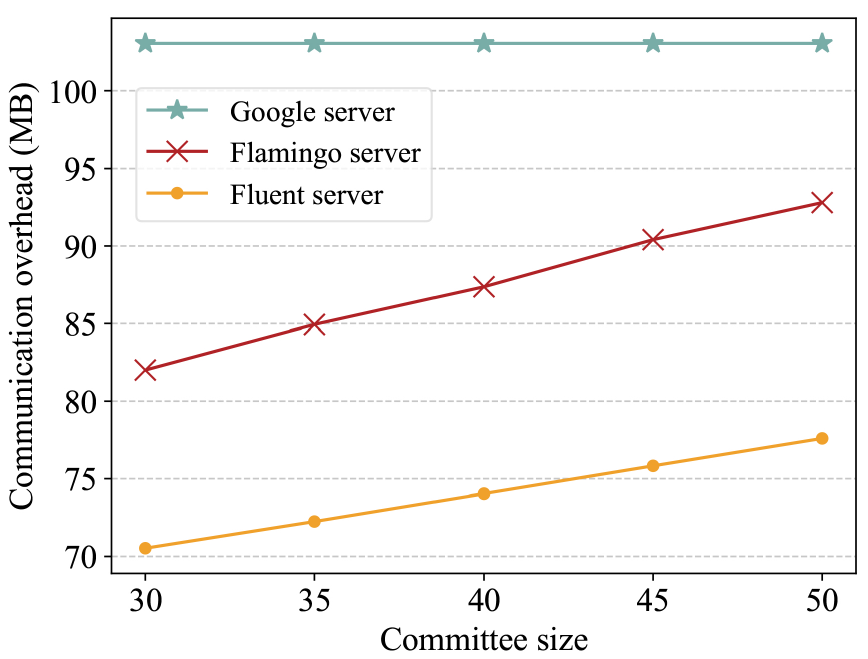}} 
      \hspace{-0.4em}
  \subfigure[Communication overhead of clients]{
    \label{committee size client communication}
    \includegraphics[height=1.25in, width=1.7in]{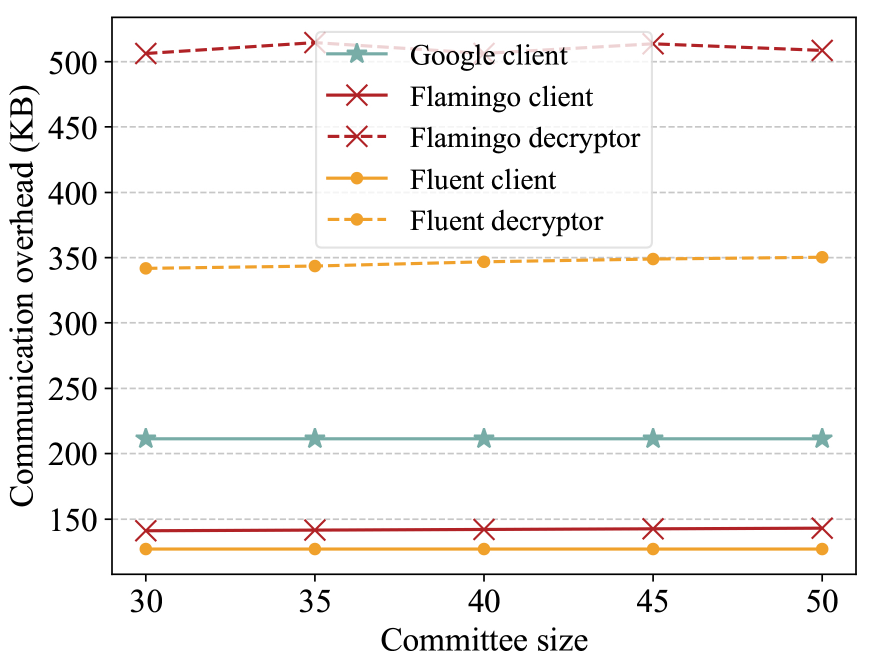}}
  \caption{Computational and communication costs in different schemes in terms of the committee size.}
  \label{committee size figure}
  }
\end{figure*}

The committee (decryptor) role assumes the crucial responsibility for mask recovery in both Flamingo and Fluent, where the size of the committee will greatly affect the computational and communication costs. Figure~\ref{committee size figure} illustrates the tendency of computational and communication costs in terms of committee size, where the size of the client is set as 500 and the dropout rate is approximately 5\%. Though the performance of Bell et al.'s scheme is irrelevant to the value of the committee size, we include it for comparative purposes.

With the increase in committee size, the computational and communication costs of servers in Flamingo and Fluent demonstrate a linear increase. This is because the consistency check in Flamingo is performed among the decryptors, and decryptors in Fluent have to compute and transmit decryption shares for other decryptors. The performance of normal clients and decryptors in both schemes remains basically the same, regardless of the size of the committee. Despite the fact that the number of neighbors in Bell et al.'s scheme is quite smaller than the number of decryptors in Flamingo and Fluent, which take on the same responsibility, the performance of Flamingo and Fluent surpasses that of Bell et al.'s scheme in terms of the communication overhead on the server side. In essence, Flamingo and Fluent aggregate more computational cost on the server side and mitigate the global communication overhead, resembling centralized machine learning, but they inherit the advantages of FL.

\subsubsection{Impact of Dropout Rate}
\begin{figure*}[!t]
  \centering{
  \subfigtopskip=2pt
  \subfigbottomskip=2pt 
  \subfigcapskip=-5pt 
  \subfigure[Computational cost of the server]{
    \label{dropout sever computation}
    \includegraphics[height=1.25in, width=1.7in]{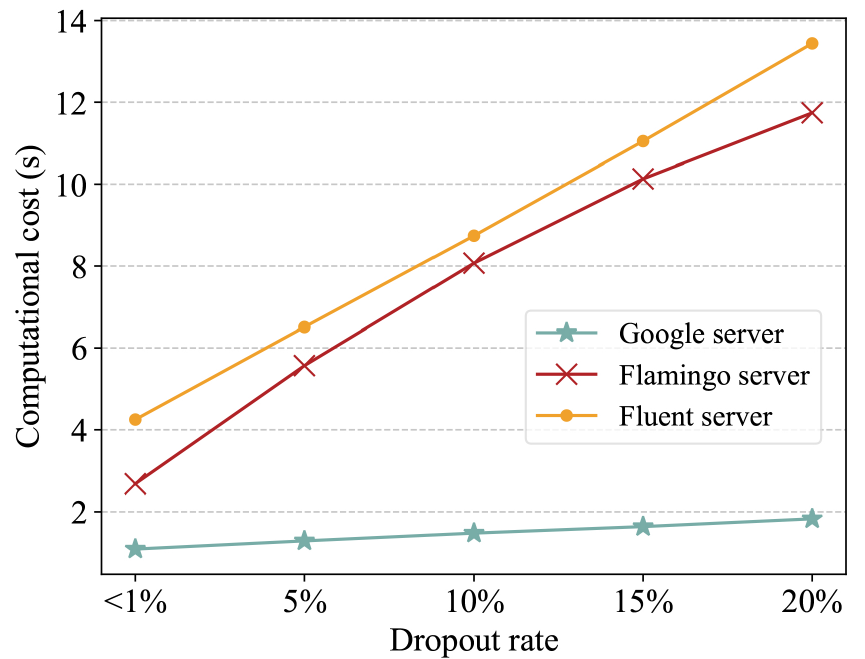}}
    \hspace{-0.4em}
  \subfigure[Computational cost of clients]{
    \label{dropout client computation}
    \includegraphics[height=1.25in, width=1.7in]{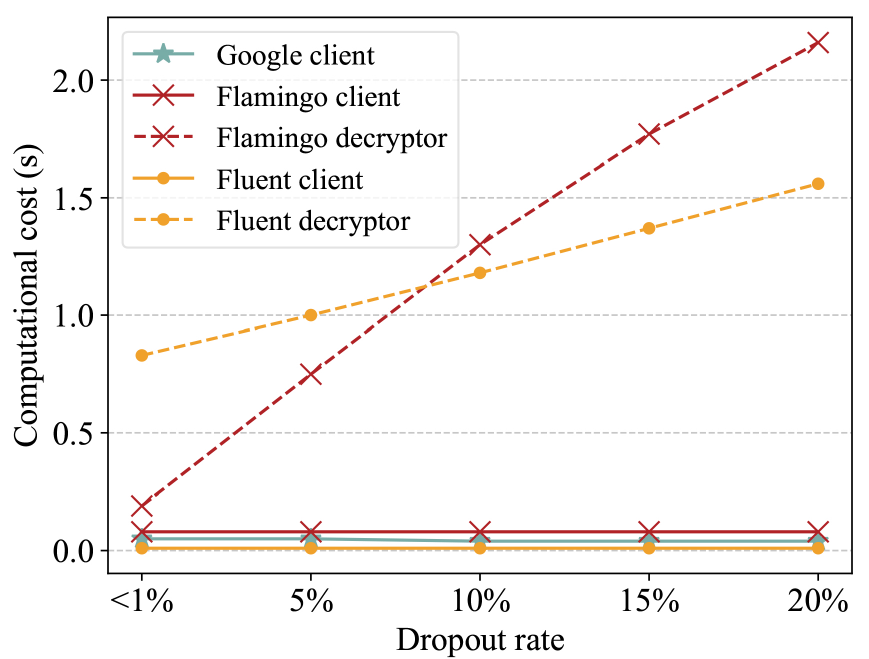}} 
    \hspace{-0.4em}
    \subfigure[Communication overhead of the server]{
    \label{dropout sever communication}
    \includegraphics[height=1.25in, width=1.7in]{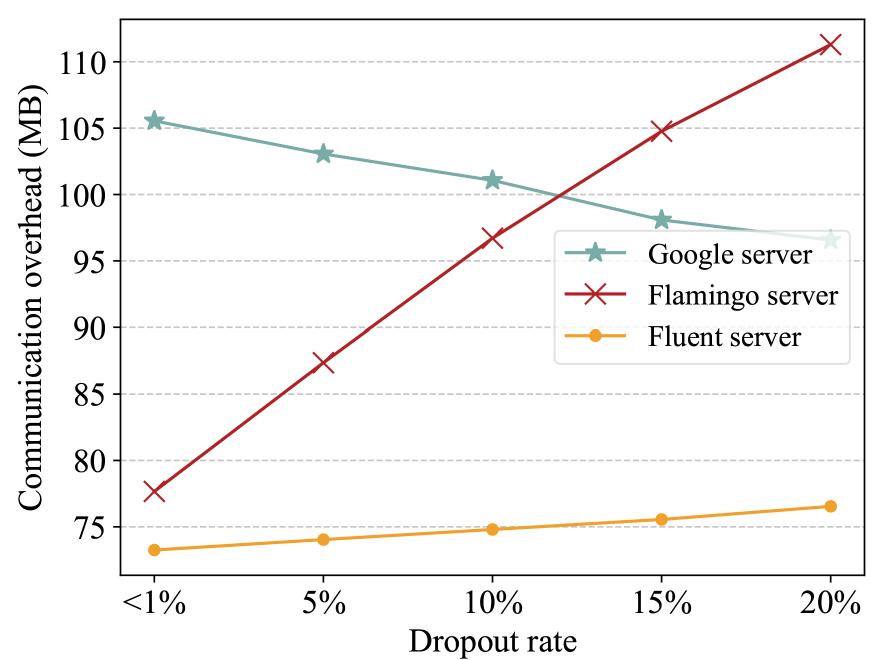}} \hspace{-0.4em}
  \subfigure[Communication overhead of clients]{
    \label{dropout client communication}
    \includegraphics[height=1.25in, width=1.7in]{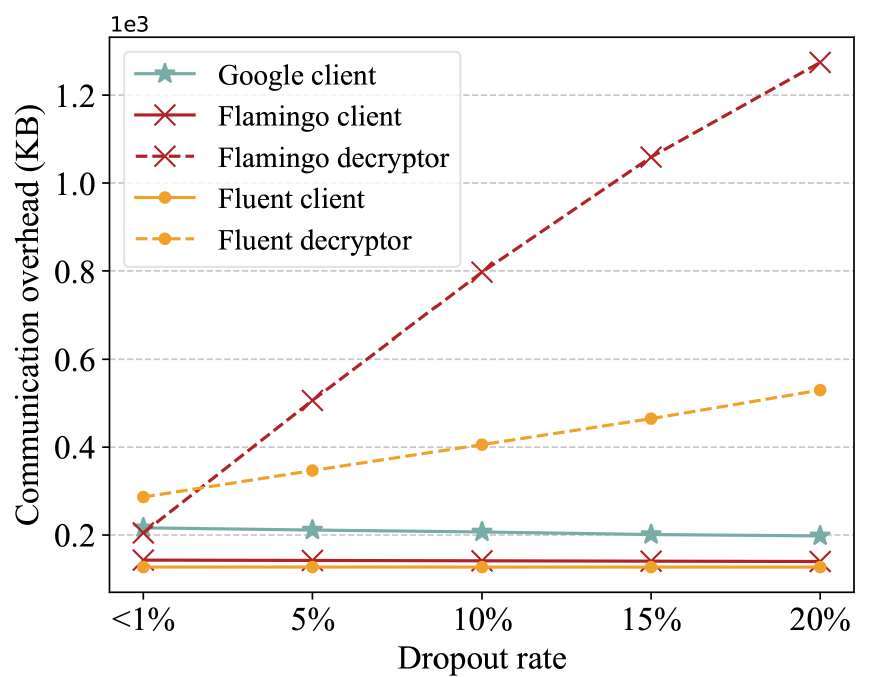}}
  \caption{Computational and communication costs in different schemes in terms of dropout rate.}
  \label{dropout figure}
  }
\end{figure*}

The dropout rate stands out as a pivotal factor influencing the performance of these schemes. Figure~\ref{dropout figure} delineates the computational and communication costs in different schemes with various dropout rates, where the client number is set as 500 and the committee size is set as 40.

Generally, the computational cost of servers and decryptors in all schemes increases linearly. The reason is that the server and the decryptors in Flamingo and Fluent have to reconstruct more seeds for the dropouts and their neighbors, facilitating the subsequent unmasking of the final vector. In Bell et al.'s scheme, clients will verify fewer signatures for consistency checks if more clients drop out. For the same reason, the communication overhead of the server and clients decreases in~\cite{BellCCS20}, while Flamingo incurs more communication overhead to transmit shares of dropouts and their neighbors. Compared to Flamingo, Fluent depicts steady expansion in terms of the computational and communication costs of the server and decryptors. This divergence in efficacy can be attributed to the demanding nature of consistency checks, imposing a substantial burden on decryptors in Flamingo. Decryptors in Flamingo bear the responsibility not only of verifying signatures from clients but also of verifying signatures from other decryptors. The dual verification process places a weighty load on decryptors, contributing to the observed performance differential. In other words, Fluent exhibits superior performance to Flamingo in dropout robustness.

Overall, Fluent exhibits comparable performance to Flamingo when confronted with fluctuations in the number of clients and committees. Meanwhile, Fluent demonstrates better efficiency in terms of the increase in dropout rate. Servers in Fluent and Flamingo~\cite{SP2023flamingo} require more computational costs than Bell et al.'s scheme~\cite{BellCCS20} because they have to execute DLP-based threshold decryption operations, which is consistent with the theoretical analysis. However, Fluent requires the least global communication overhead because of the one-round consistency check and unmasking design. Clients in Fluent only need around 0.01s and 127 KB to compute and transmit the masked vector, regardless of the number of clients, committee size, and dropout rate. This is because normal clients have shared secret self masks and pairwise masks with other clients or decryptors in the pre-round setup phase. This characteristic is a significant advantage since any resource-limited device can also bear the resource requirement. It is worth noting that the extra computational cost of the decryptors and the server reduces the communication rounds from 3 rounds in state-of-the-art schemes to 2 rounds in Fluent, which is especially important for geographically distributed clients.

\subsection{Performance in FL}
We demonstrate the effectiveness of Fluent with two practical use cases in FL. In particular, we train a convolutional neural network (CNN) on both the MNIST and CIFAR-10 datasets from Torchvision. Both implements are realized among 100 clients with 5 local epochs, 10 local batch sizes, and a learning rate $\gamma=0.01$. The same network delay model as Flamingo~\cite{SP2023flamingo} is leveraged, where a jitter is added over base delay with default parameters in ABIDES. Table~\ref{table: dataset training result} shows the experimental results, where a model architecture typed 2xConv2D-MaxPooling-Flatten-2xDense for MNIST and a model architecture typed 2x(Conv2D-MaxPooling)-Flatten-3xDense for CIFAR-10 with PyTorch are constructed and trained for 100 and 300 iterations, respectively.

Overall, Fluent requires a longer waiting time for decryptors to compute encrypted shares, resulting in slower model training. However, Fluent requires the least computational and communication costs for normal clients and offloads more computational costs to the server and decryptors. Note that we simulate all agents with the same PC, the performance of the server and decryptors may be better in practice. The schemes may be suitable for scenarios where both lightweight and heavyweight devices exist and powerful institutions or devices play the role of decryptors.

\renewcommand\arraystretch{1.3}
\begin{table*}[htbp]
\caption{The concrete training time, computational and communication costs for specific federated learning applications, where ``x / y'' represents the computational and communication costs of normal clients and decryptors, respectively.}
\begin{center}
\resizebox*{\linewidth}{!}{
\begin{tabular}{cccccccccc}
\hline
Dataset & Scheme   &  \makecell{Training\\time} & \makecell{computational cost\\ of server (s)} & \makecell{computational cost\\ of client (s)} & \makecell{communication\\ of server (MB)} & \makecell{communication\\ of client (MB)}  \\ 
\hline
\multirow{3}{*}{\makecell{MNIST\\(21840 parameters,\\100 iterations)}} & \cite{BellCCS20} & 1h 49min & 14.34 & 226.58 &  2224.14 & 22.26 \\
& Flamingo \cite{SP2023flamingo}  & 52min & 132.31 &  228.20 / 4.85 & 1918.23 & 17.86 / 4.53 \\
& Fluent   & 1h 25min & 95.19 & 223.64 / 18.65 & 1909.44 & 16.87 / 7.43 \\
\hline
\multirow{3}{*}{\makecell{CIFAR-10\\(62006 parameters,\\300 iterations)}} & \cite{BellCCS20}  &  5h 27min & 50.31 & 396.69 & 15720.69 & 157.28 \\
& Flamingo \cite{SP2023flamingo} &  2h 35min & 318.93 & 401.78 / 13.65 & 14933.62 & 145.51 / 13.01 \\
& Fluent  & 4h 15min & 308.09 & 389.48 / 55.85 & 14920.80 & 142.55 / 22.27 \\
\hline
\end{tabular}
}
\label{table: dataset training result}
\end{center}
\end{table*}

\section{Extension and Discussion} \label{extension}
\subsection{Fluent-Dynamic: Fluent with Dynamically Joining Participants}
In Flamingo, a novel graph generation algorithm is proposed for the server and clients to locally generate neighbors and the global graph of participants. However, Flamingo fixes the set of all clients ($\mathcal{N}$) involved in a training session and the number of clients $n_t$ for iteration $t$. The static arrangement impedes the dynamic inclusion of clients during iterations, imposing significant constraints on the flexibility and scalability of scenarios and applications. Therefore, we try to extend Fluent with the capability of dynamic participant joining. In this way, Fluent will have similar functionality to Bell et al.'s scheme~\cite{BellCCS20} but with fewer communication rounds.

It is reasonable to adapt the size of the committees to accommodate the arrival of new clients. Since all clients share self and pairwise seeds with decryptors, decryptors in Fluent own more shares of seeds than those in Flamingo. Therefore, it is expensive to utilize the same method as in Flamingo to re-share the shares of decryptors. Additionally, their approach necessitates all existing decryptors to re-share their seeds, resulting in a significant waste of resources, particularly when only clients join the committee.

As above, we will provide two corresponding modifications that are compatible with Fluent and further enable Fluent to satisfy the requirement of dynamic participant joining.

\noindent\textbf{Modification 1.}
To solve the first problem, a new participant selection scheme is devised with the idea in~\cite{CCS2021FMD}, i.e., each client $i$ has a probability $p=n/m$ of being selected in each iteration instead of a fixed number $n_t$. Specifically, we map the concatenation of global model $\Theta_t$, iteration $t$, and the index of client $i$ to a $\lambda$-bit random number $\alpha$. Then, convert $\alpha$ to a base-$m$ number and get its last digital $\beta$. Those numbers whose last digit is smaller than $n$ will be selected as participants in iteration $t$. The new participant selection algorithm is shown in Figure~\ref{participant slection}. Provided that the hash function $H$ is completely random, each client has a fair probability of being selected in each iteration. Although the number of participants for each iteration may be dynamic, the assumption guarantees that the deviation will be small enough to preserve the privacy of clients.

\begin{figure}[htbp]
\centering
\framebox{%
\begin{minipage}[htbp]{0.45\textwidth}
\makebox[5em][l]{\textbf{Parameters:} Security parameter $\lambda$, global model $\Theta_t$ of} \\
\makebox[5em][l]{$\qquad$iteration $t$, a hash function $H:\{0,1\}^*\rightarrow \{0,1\}^{\lambda}$,} \\
\makebox[5em][l]{$\qquad$probability $p=n/m$, and the number of clients $N$.}\\
\makebox[5em][l]{\textbf{Output:} $S_t$.} \\
\makebox[5em][l]{\textbf{function} ChooseSet$(\Theta_t,t, p, N)$} \\
\makebox[5em][l]{$\quad S_t\leftarrow \emptyset.$} \\
\makebox[5em][l]{$\quad$For i in $[N]:$} \\
\makebox[5em][l]{$\quad\quad\alpha=H(\Theta_t\parallel t \parallel i)$.} \\
\makebox[5em][l]{$\qquad$Convert $\alpha$ to a base-$m$ number and get last digit $\beta$.} \\
\makebox[5em][l]{$\qquad$\textbf{if} $\beta < n$:} \\
\makebox[5em][l]{$\qquad\quad$Add $i$ into $S_t$.} \\
\makebox[5em][l]{$\quad$Output $S_t$.} 
\end{minipage}
}
\caption{New participant selection algorithm.}
\label{participant slection}
\end{figure}

\noindent\textbf{Modification 2.} 
To solve the second problem, we utilize multilevel (hierarchical) threshold secret sharing~\cite{MTSS1998,HTSS2007JoC,HarnM14MTSS} to replace Shamir's threshold secret sharing scheme. Specifically, a set $\mathcal{P}=\{\mathcal{P}_1,\mathcal{P}_2,\dots,\mathcal{P}_n\}$ of $n$ participants is partitioned into $l$ disjoint subsets $\mathcal{P}_1, \mathcal{P}_2, \dots, \mathcal{P}_l$. The subset $\mathcal{P}_1$ is on the highest level, while $\mathcal{P}_l$ is on the least privileged level. Denote the number of participants on the $i$-th level as $n_i=\lvert \mathcal{P}_i \rvert$. The threshold $t_i$ indicates the smallest number of participants on the $i$-th or higher levels who can cooperate to successfully reconstruct the secret. Let $N_i$ be the total number of participants on the $i$-th and higher levels, i.e., $N_i=\sum_{j=1}^i n_j,1\leq i\leq l$. Generally, we assume that the thresholds at different levels satisfy the relation $\kappa_1<\kappa_2<\dots<\kappa_l$. The access structure is defined as:
\begin{equation*}
    \Gamma = \{\mathcal{A}\subseteq \mathcal{P} \mid \sum_{j=1}^i \lvert \mathcal{A}\cap \mathcal{P}_j \rvert \geq \kappa_i~\mathrm{for}~i=1,2,\dots,l\}.
\end{equation*}

Among the existing multilevel (hierarchical) threshold secret sharing schemes that have been proposed~\cite{MTSS1998,HTSS2007JoC,HarnM14MTSS}. Ghodosi et al.'s scheme~\cite{MTSS1998} is selected for Fluent since it is compatible with the privacy preservation approach in Fluent. In their scheme, each shareholder has only one share of the secret distributed by the dealer. To construct the scheme, a dealer first utilizes Shamir's $(\kappa_1, N_1)$ secret sharing, i.e., constructs a polynomial $f_1(x)=K+a_{1,1}x+\dots+a_{1,T_1}x^{T_1}$, where $T_1=\kappa_1-1$ and distributes shares of $K$ to set $\mathcal{P}_1$. Then, the dealer extends the scheme to a $(\kappa_2,N_2)$ scheme for the next set $\mathcal{P}_2$ in the next level. The dealer will reconstruct a new polynomial $f_2(x)=K+a_{2,1}x+\dots+a_{2,T_2}x^{T_2}$ satisfying $f_2(i)=f_1(i)$ for $i\in \mathcal{P}_1$ and calculate new shares of the secret $K$ for $\mathcal{P}_2$ with $f_2$. The rest may be deduced by analogy. In this way, newly added decryptors can also obtain secret shares of the same self and pairwise masks without re-sharing the seeds with existing decryptors. Note that version control is necessary in the proposed scheme to tag the index and level of decryptors and prevent frequent polynomial construction.

\noindent\textbf{Advantages.} Compared with re-executing the pre-round phase for all clients, including newly added clients, the enhanced scheme obviates the necessity for another handshake for key negotiation or secret sharing among existing clients and decryptors. In other words, newly added clients execute the whole pre-round phase, whereas existing clients merely necessitate handshakes with newly added clients and share the new pairwise masks with decryptors. Similarly, upon the inclusion of new decryptors into the committee, the existing decryptors are likewise exempted from the necessity of updating their private information. Considering that the number of new clients will be far fewer than existing clients, our extension promises a substantial reduction in both computational cost and communication overhead, especially when there have been tremendous existing clients or decryptors.

\noindent\textbf{Security.} When new clients join the training process, it is imperative for them to transmit their public keys for commitment with a Merkle tree. The Merkle tree provides a verifiable mechanism for clients to validate the public keys of all participants. In other words, the security of the proposed scheme can be guaranteed even if the server is not completely honest during the public key commitment. When new decryptors join the committee, the shares of self and pairwise masks are shared among more decryptors. Nevertheless, the thresholds for secret reconstruction were also updated for clients at different levels. Consequently, the security of multilevel threshold secret sharing guarantees the prevention of any privacy information leakage in this particular scenario. In summary, the dynamic inclusion of clients and decryptors is practical and does not introduce additional security vulnerabilities.

\subsection{An Alternative Cross-checking Approach} \label{section: cross-checking}
The consistency check and unmasking approach introduced in \S \ref{detail scheme} reduces one round of communication in the collection phase. However, the computational cost of the server and decryptors increases to generate decryption key shares and mask reconstruction. Alternatively, we introduce another consistency check design as an alternative approach for different scenarios and applications. The approach has the same communication round and similar computational cost as existing schemes~\cite{BellCCS20, SP2023flamingo} but decreases the communication of one step from $\mathcal{O}({n_I}^2)$ to $\mathcal{O}(1)$, where ${n_I}$ is the number of decryptors.

The intuition is that the server and decryptors may collaboratively generate a signature on the set of survivors and dropouts with a non-interactive threshold signature scheme~\cite{BoldyrevaPKC03TSS}. Then, the signature will be transmitted to the decryptors and verified with the master public key to improve communication efficiency. According to the security of the threshold signature scheme~\cite{BoldyrevaPKC03TSS,BachoBLSCCS22} (see Appendix~\ref{threshold signature scheme}), a complete signature can be generated only when at least threshold $\kappa$ decryptors sign the set of survivors and dropouts. In this way, the server does not need to transmit the signatures of all decryptors for consistency checks.

Specifically, after collecting enough vectors and signatures for a determined time period, the server defines a global set of dropouts $\mathcal{U}_\mathcal{D}$ and a set of survivors $\mathcal{U}_\mathcal{S}$. Then, the server transmits the signatures of the survivors to the decryptors for verification. Afterward, if all signatures are legitimate and the sets are correct, the decryptors compute partial signatures $\sigma_{\mathcal{U}_\mathcal{D}}^{(i)}\leftarrow$ TSS.Sign$(msk_i,\mathcal{U}_\mathcal{D})$ and $\sigma_{\mathcal{U}_\mathcal{S}}^{(i)}\leftarrow$ TSS.Sign$(msk_i,\mathcal{U}_\mathcal{S})$ on $\mathcal{U}_\mathcal{D}$ and $\mathcal{U}_\mathcal{S}$ with a threshold signature scheme and secret shares of the master secret key $msk$. Then, decryptors will return the partial signatures to the server. The server aborts if it receives fewer than $(1-\eta)n_t$ responses. Otherwise, the server aggregates all partial signatures and obtains complete signatures $\sigma_{\mathcal{U}_\mathcal{D}}$ and $\sigma_{\mathcal{U}_\mathcal{S}}$. It sends the aggregated signatures $(\mathcal{U}_\mathcal{D}, \sigma_{\mathcal{U}_\mathcal{D}})$ and $(\mathcal{U}_\mathcal{S}, \sigma_{\mathcal{U}_\mathcal{S}})$ to all decryptors for legitimacy verification. The decryptors verify the signatures with $mpk$ and abort if any of them fails to be verified.

\section{Related Work} \label{related works}

Many PPFL schemes and systems employing various cryptographic primitives have been proposed, e.g. differential privacy (DP), multi-party computation (MPC), Homomorphic Encryption (HE), etc.

\noindent\textbf{Masking mechanism.} Some works are dedicated to decreasing the communication round of secure aggregation based on masking mechanisms~\cite{SP2023flamingo, ASIACRYPT2023LERNA, MicroFedML}. However, all of them need three communication rounds in the malicious server setting. Flamingo only needs to share self masks in each iteration, and the distributed generated keys for decryptors are reusable. The idea reduces one communication round by utilizing threshold decryption to re-use decryption key shares. However, the computational and communication costs for self mask sharing and reconstruction, or masking encryption and decryption, are still required. Different from Flamingo, LERNA~\cite{ASIACRYPT2023LERNA} and MicroFedML~\cite{MicroFedML} are based on the single-masking foundation and try to reuse the pairwise masks by different solutions. LERNA~\cite{ASIACRYPT2023LERNA} utilizes key-homomorphism that relies on different mathematical structures (underlying the LWR and DCR assumptions) to obtain the aggregated sum. MicroFedML~\cite{MicroFedML} utilizes the hardness of DLP to aggregate and exclude dropouts for privacy preservation. However, their scheme has to brute force the DLP to calculate the summary of gradients during aggregation, which is feasible only if the summary of gradients is sufficiently small~\cite{ShiNDSS2011}. Besides, the characteristic that most gradient entries are decimal numbers with long decimal digits also increases difficulty in resolving the issue. Therefore, their scheme is applicable in gradient sparsification, quantization, and weight regularization areas in FL. In contrast, Fluent does not have any restrictions on the input domain.

\noindent\textbf{Differential privacy (DP).} DP techniques theoretically protect privacy over multiple FL rounds by adding statistical noise~\cite{IJCAI2021LDP, USENIX2022DP, INFORCOM2023DP, CCS2023Xie}. The main concern with the solution is that added noise comes at the expense of the accuracy of the aggregated model~\cite{AAAI2023DPSun, USENIX2023DPaccuracy}. It is worth noting that secure aggregation and DP are complementary, i.e., the benefits of DP can be applied to the secure aggregation protocols by adding noise to the local models. A recent work~\cite{USENIX2022DP} combines differential private technology and masking structure to perform secure aggregation. They instantiate their protocol with a learning with errors (LWE)-based algorithm, where the noise added for differential privacy also serves as the noise term in LWE.

\noindent\textbf{Multi-party computation (MPC).} The framework of two or more non-colluding servers~\cite{PETs2023} is common to achieve distributed trust and privacy preservation~\cite{PrioNSDI2017, Prio+SCN2022, ELSASP2023, IACR2023ScionFL}. A promising approach named ELSA~\cite{ELSASP2023} leverages two servers to aggregate vectors with the technologies of arithmetic/boolean secret sharing and oblivious transfer (OT). Furthermore, their scheme can withstand model poisoning attacks~\cite{CCS2023MESAS} by detecting and filtering out boosted gradients with respect to Euclidean distance. Distributing equal-length vectors to aggregators and the interaction between aggregators cause high computational and communication costs, which is a main problem to be solved.

\noindent\textbf{Other works.} Some other protocols use \textbf{trusted execution environments (TEE)}~\cite{mobisys2021PPFL, SHUFFLEFLCF2021,TDSC2022TEE} as the backbone of PPFL, e.g. ARM TrustZone and Intel Software Guard Extensions (SGX). Though TEE guarantees confidentiality and integrity of the code and data, the limited physical trusted memory (128 MB for current Intel CPUs) restricts its large-scale applications, especially when the vectors of clients have high dimensions. \textbf{Homomorphic encryption (HE)}~\cite{ShiNDSS2011, usenixATC2020HE, tian2021secure, TDSC2023HE, choffrut2023sable} is also widely adopted owing to its computability over ciphertexts. However, the extensive computational demands of HE, particularly for high-dimensional ML models, have deterred attempts to design purely homomorphic operators for non-linear robust aggregators~\cite{choffrut2023sable}. \textbf{Secure shuffling}~\cite{SHUFFLEFLCF2021, CCS2023shuffling} is another approach to PPFL that relies on anonymous communication assumptions. However, it is challenging to have completely anonymous communication channels between the clients and the server.

\section{Conclusion and Future work} \label{conclusion}
We present Fluent, a new construction that achieves the least communication round in masking-based secure aggregation schemes for FL. The construction re-uses the shares of masks and realizes an one-round consistency check and unmasking. Formal security proof demonstrates the security of Fluent in the malicious server setting. Experiments demonstrate the superiority of Fluent over state-of-the-art schemes in terms of efficiency and practicality. In addition, Fluent can be extended to support dynamic client and committee joining, which enhances its flexibility and scalability.

As future work, we aim to consider more security threats in FL. For instance, defenses against poisoning attacks (untargeted poisoning attacks and targeted backdoor attacks) are important to guarantee the integrity and correctness of the global model. How to prevent such attacks is challenging due to their stealthy nature.

\bibliographystyle{plain}
\bibliography{mybibfile.bib}


\appendix

\section{The Algorithm of Fluent} \label{appendix: scheme}
A detailed algorithm of Fluent is shown in Figure~\ref{fig: main algorithms}.

\begin{figure*}[htbp]
\centering
\framebox{
\begin{minipage}[htbp]{0.96\textwidth}
\textbf{Public parameters:} Security parameter $\lambda$, a threshold $\kappa$, a set of clients $\mathcal{N}$, a Merkle tree $\mathcal{T}$, input domain $\mathbb{X}$, vector length $l$, a pseudorandom generator $PRG$: $\mathbb{Z}_q \rightarrow \mathbb{X}^l$, and a map-to-point hash $H_{mtp}$: $\{0,1\}^*\rightarrow \mathbb{G}$. \\
\textbf{Client $i$'s input:} $\mathbf{x_i}\in \mathbb{X}^l$  \\
\textbf{Output:} $\mathbf{z} \in \mathbb{X}$ 

\textbf{Pre-round phase} 
\begin{compactenum}
    \item[1.] Each client $i$ generates three key pairs $(sk_{i,1},pk_{i,1}=g^{sk_{i,1}}),(sk_{i,2},pk_{i,2}=g^{sk_{i,2}}),(sk_{i,3},pk_{i,3}=g^{sk_{i,3}})$ and sends $(pk_{i,1},pk_{i,2},pk_{i,3})$ to the server.

    \item[2.] The server commits to the public key vectors $pk_1=(pk_{i,1})_i,pk_2=(pk_{i,2})_i,pk_3=(pk_{i,3})_i$ for $i\in \mathcal{N}$ with a Merkle tree. It sends the signed root hashes $h_{root,1}$, $h_{root,2}$, and $h_{root,3}$ to each client.

    \item[3.] Selects a random set of decryptors $\mathcal{I}\leftarrow \mathrm{ChooseSet}(h_{root,1}\parallel h_{root,2} \parallel h_{root,3}, t, n_{\mathcal{I}}, \mathcal{N})$.
    
    \item[4.] With a DKG protocol, all decryptors and the server hold the master public key $mpk=g^{msk}$, and each decryptor holds a secret share $msk_i\in \mathbb{Z}_q$ of the master secret key $msk$.

    \item[5.] Every two clients $i,j\in \mathcal{N}$ negotiate a secret seed $seed_{i,j}=\mathrm{KA.Agree}(sk_{i,1},pk_{j,1})$. 

    \item[6.] Each client with identity $i$ selects a random number $seed_i\in \mathbb{Z}_q$ and generates two sets of shares $(u,seed_i^{(u)})_{u\in \mathcal{I}}\leftarrow \mathrm{SS.Share}(seed_i,\kappa,\mathcal{I})$ and $(u,seed_{i,j}^{(u)})_{u\in \mathcal{I}}\leftarrow \mathrm{SS.Share}(seed_{i,j},\kappa,\mathcal{I})$.

    \item[7.] For $u\in \mathcal{I}$, each client with identity $i$ computes session key $key_{i,u}=\mathrm{KA.Agree}(sk_{i,2},pk_{u,2})$ and $c_{i,u} \leftarrow \mathrm{AE.Enc}(key_{i,u}, i\parallel u \parallel seed_i^{(u)} \parallel seed_{i,j}^{(u)})$ and sends $(u,c_{i,u})$ to client $u$ via the server.

    \item[8.] Client $u$ decrypts the ciphertext $c_{i,u}$ with $k_{u,i}$ and obtains $\{i\parallel u \parallel seed_i^{(u)} \parallel seed_{i,j}^{(u)}\}$. Store the message if the index is correct.
\end{compactenum}

\textbf{Collection phase} 

\textbf{1. Report} 
\begin{compactenum}
    \item[(1)] The server computes $S_t\leftarrow \mathrm{ChooseSet}(\Theta_t, t, n_t,\mathcal{N})$ and sends the global model $\Theta_t$ to be updated to each client $i\in S_t$.
    
    \item[(2)] Then, each client that receives the message computes $S_t\leftarrow \mathrm{ChooseSet}(\Theta_t, t, n_t,\mathcal{N})$ and $A_{i,t} \leftarrow \mathrm{FindNeighbor}(\Theta_t,S_t,i)$ locally to define the neighbors of $i$ as $A_{i,t}$.

    \item[(3)] Each client $i\in S_t$ that has not dropped out performs the following:
        \begin{compactitem}
            \item Trains the model with its local database $D_i$ and obtains an update $\mathbf{x_i}$.
            \item Computes $g_t=H_{mtp}(\Theta_t,t)$, $\mathbf{r_i}=PRG(g_t^{seed_i})$, $\mathbf{m_{i,j}}=PRG(g_t^{seed_{i,j}})$, and $\mathbf{sk_i}=\mathbf{r_i}-\sum_{0<j<i}\mathbf{m_{i,j}}+\sum_{i<j\leq n_t}\mathbf{m_{i,j}}$.
            \item Computes the masked input as $\mathbf{y_i}=\mathbf{x_i}+\mathbf{sk_i}$.
            \item Computes $\sigma_{i}\leftarrow \mathrm{DS.Sign}(sk_{i,2},m_i=``online'' \parallel i \parallel t)$.
            \item Send the masked gradient $\mathbf{y_i}$ and the message and signature pair $(m_i,\sigma_i)$ to the server.
        \end{compactitem}
\end{compactenum}

\textbf{2. Consistency check and unmasking} 
\begin{compactenum}
    \item[(4)] The server collects messages for a determined time period. If it receives fewer than $\kappa$ messages, it aborts. Otherwise, it checks the legitimacy of signatures and defines a global set of dropouts $\mathcal{U}_\mathcal{D}$ and a set of survivors $\mathcal{U}_\mathcal{S}$. It then sends the message and signature pairs $(m_i,\sigma_{i})$ to every decryptor $u\in \mathcal{I}$.

    \item[(5)]The decryptor $u$ performs as follows:
        \begin{compactitem}
            \item Checks that $\mathcal{U}_\mathcal{S}\cap \mathcal{U}_\mathcal{D} = \emptyset$, $\lvert \mathcal{U}_\mathcal{S} \rvert \geq (1-\eta)n_t$, and that all signatures $\sigma_{i}$ on message $m_{i}$ for all $i \in \mathcal{S}_t$ are valid, aborting if any of these checks fails.
            \item Computes $g_t=H_{mtp}(\Theta_t,t)$, $k_u\xleftarrow{\$}\mathbb{Z}_q $, $c_{seed,u}=\mathrm{AE.Enc}(k_u,g_{t}^{seed_i^{(u)}}\parallel g_{t}^{seed_{j,k}^{(u)}})_{i\in \mathcal{U}_\mathcal{S},j\in \mathcal{U}_\mathcal{D},k\in A_{j,t}}$, $c_{key,u}=Encode(k_u)\cdot mpk^{sk_{u,3}+H(\mathcal{U}_\mathcal{S}\parallel \mathcal{U}_\mathcal{D})}$, where $Encode$ encodes $k_u$ to an element of $\mathbb{G}$.
            \item Computes $c_{u,i}={(g^{H(\mathcal{U}_\mathcal{S} \parallel \mathcal{U}_\mathcal{D})}\cdot pk_{i,3})}^{msk_u}$ for $i\in \mathcal{I} \backslash \{u\}$.
            \item Sends the message $\{c_{seed,u},c_{key,u},c_{u,i}\}_{i\in \mathcal{I} \backslash \{u\}}$ to the server.
        \end{compactitem}

    \item[(6)] The server performs as follows:
    \begin{compactitem}
        \item Aborts if it receives fewer than $\kappa$ responses.
        \item Computes $k_u=Decode(c_{key,u}/\Pi_{i\in \mathcal{I} \backslash \{u\}}{c_{i,u}}^{\beta_i})$ and $\{g_{t}^{seed_i^{(u)}}\parallel g_{t}^{seed_{j,k}^{(u)}}\}_{i\in \mathcal{U}_\mathcal{S},j\in \mathcal{U}_\mathcal{D}, k\in A_{j,t}} = \mathrm{AE.Dec}(k_u,c_{seed,u})$, where $\beta_i$ represents the Lagrange coefficients.
        \item For $u\in \mathcal{I}$, computes $g_t^{seed_i}=\Pi_{u=1}^{\kappa}({g_{t}^{seed_i^{(u)}}})^{\beta_u}$ and $g_{t}^{seed_{j,k}}=\Pi_{u=1}^{\kappa}({g_{t}^{seed_{j,k}^{(u)}}})^{\beta_u}$.
        \item Unmasks the private vectors and obtains the aggregated model as $\mathbf{z}=\sum_{i\in \mathcal{U}_\mathcal{S}} \mathbf{x_i}=\sum_{i\in \mathcal{U}_\mathcal{S}}\mathbf{y_i}-\mathbf{sk}$, where $\mathbf{sk}=\sum_{i\in \mathcal{U}_\mathcal{S}}PRG(g_t^{seed_i})\pm \sum_{j\in \mathcal{U}_\mathcal{D}, k\in A_{j,t}} PRG(g_{t}^{seed_{j,k}})$.
    \end{compactitem}
\end{compactenum}
\end{minipage}
}
\caption{Round-efficient secure aggregation for long-term FL.}
\label{fig: main algorithms}
\end{figure*}

\section{Security Proof} \label{appendix: security proof}

\subsection{Proof of Theorem 1} \label{proof: theorem 1}
Due to the security of Shamir's secret sharing scheme, the PPT adversary cannot reconstruct $g_t^{sk}$ from any set $\{g_t^{sk_u}\}_{u\in [\kappa-1]}$, where $t\in [t_1]$. Therefore, the $g_t^{sk}$ and $g_t^{sk_u}$ essentially share the same security. For simplicity, we prove that all $\{g_t^{sk}\}_{t\in[t_1]}$ will not leak the privacy of subsequent $\{g_{t'}^{sk}\}_{t_1<t'\leq T}$ and the same holds for $g_t^{sk_u}$.

Obviously, there are the following routes to compute ${sk}$ or $g_t^{sk}$, and we will analyze their privacy preservation as follows:
\begin{itemize}[leftmargin=*]
    \item Given $\{g_t^{sk}\}_{t\in [t_1]}$ and $g_t$, it is computationally intractable to compute $sk$ due to the DL assumption.
    \item Given $\{g_t^{sk}\}_{t\in [t_1]}$ and $g_t$, it is computationally intractable to compute $\{g_{t'}^{sk}\}_{t_1<t'<T}$ due to the assumption of CDH. Specifically, $g_{t'}$ can be seen as $g_t^{r}$, and it is computationally intractable to compute $g_t^{r\cdot sk}=g_{t'}^{sk}$.
\end{itemize} 
\hfill   $\square$

\subsection{Proof of Theorem 2} \label{proof: theorem 2}
We demonstrate that our protocol is a secure summation protocol in a malicious server setting. We prove the theorem statement by defining a simulator Sim through a sequence of hybrids so that the views of the adversary $\mathcal{A}$ in any two subsequent executions are computationally indistinguishable. According to the defined threat model, the adversary holds control over the server and a fraction of the clients and does not interact with the honest clients, but with the simulator Sim that does not have access to the inputs of honest parties. We prove that the joint view of the adversary reveals no information about the inputs of other clients beyond what can be deduced from the output of the computation. We also assume that the simulator Sim can query once an oracle computes an ideal functionality that captures the leakage that we are willing to tolerate. In each of the hybrids below, to ease notation, we assume that the Sim will cause honest parties to abort as they would during the real protocol execution (e.g., when receiving a malformed message). We denote $\mathcal{H}=S_t\backslash \mathcal{U_C} \backslash \mathcal{U_D}$ the set of honest clients.

\begin{enumerate}
    \setlength{\itemindent}{1em} 
    \item[\textbf{Hyb$_1$.}] This is the real execution of the protocol, where the adversary is interacting with honest parties.

    \item[\textbf{Hyb$_2$.}] In this hybrid, we introduce a simulator Sim, which knows the private vector $\mathbf{x_i}$, the shares of master secret key $msk_i$, self seed $seed_i$, private keys $sk_{i,1},sk_{i,2},sk_{i,3}$ for all honest parties $i\in \mathcal{H}$. The view of the adversary $\mathcal{A}$ in this hybrid is the same as in the previous hybrid.
    
    \item[\textbf{Hyb$_3$.}] In this hybrid, Sim aborts any of the honest parties in $\mathcal{H}$ if $\mathcal{A}$ provides a public key for $i\in \mathcal{H}$ which is not the valid public key of the client $i$. The view of the adversary $\mathcal{A}$ in this hybrid is indistinguishable from the view in the previous hybrid, since we assumed the server to behave honestly during the public key commitments phase with a Merkle tree, and each client $i\in \mathcal{H}$ checks the legitimacy of the keys it received from Sim are in the Merkle tree.

    \item[\textbf{Hyb$_4$.}] In this hybrid, Sim replaces the shares of the actual master secret key $msk$ sent to the honest client $i\in \mathcal{H}$ with the shares of zeros (but still reveals the real shares in Step (6)). Since $\lvert \mathcal{U_C} \rvert \leq \eta_\mathcal{C} \lvert S_t \rvert$, $\mathcal{A}$ cannot receive sufficient shares to reconstruct $msk$. Therefore, the information-theoretic security of Shamir's secret sharing guarantees that the view of the adversary $\mathcal{A}$ in this hybrid is indistinguishable from the view in the previous hybrid.
    
    \item[\textbf{Hyb$_5$.}] In this hybrid, Sim replaces $\mathrm{KA.Agree}$ in Step 5 by $\mathrm{Sim}_{KA}(seed_{i,j})$, where $seed_{i,j}$ is chosen uniformly at random for all $i,j\in \mathcal{H}$. The view of the adversary in this hybrid is indistinguishable from the view in the previous hybrid because of the security of the key agreement protocol.
    
    \item[\textbf{Hyb$_6$.}] In this hybrid, Sim replaces $\mathrm{KA.Agree}$ in Step 7 by $\mathrm{Sim}_{KA}(key_{i,j})$, where $key_{i,j}$ is chosen uniformly at random for all $i,j\in \mathcal{H}$. The view of the adversary in this hybrid is indistinguishable from the view in the previous hybrid because of the security of the key agreement protocol.

    \item[\textbf{Hyb$_7$.}] In this hybrid, Sim aborts the client $i\in \mathcal{H}$ if $i$ receives from the server a ciphertext $(j,c_{i,j})$ for $j\in \mathcal{H}$ which is not the ciphertext created by $i$ and can be successfully decrypted. Since the key $key_{i,j}$ was chosen uniformly at random in Hyb$_3$, the view of the adversary in this hybrid is indistinguishable from the view of the previous hybrid by the IND-CTXT security of the AE scheme.

    \item[\textbf{Hyb$_8$.}] In this hybrid, Sim replaces all encrypted shares sent between $i,j\in \mathcal{H}$ in Step 7 with encryptions of zeros (but still reveals the real shares in Step (6)). Since the key $key_{i,j}$ was chosen uniformly at random in Hyb$_6$, the view of the adversary $\mathcal{A}$ in this hybrid is indistinguishable from the view in the previous hybrid by the semantic security of the AE scheme.

    \item[\textbf{Hyb$_9$.}] In this hybrid, Sim aborts client $i\in \mathcal{H}$ in Step (3) if any of the signatures $\sigma_{i}$ for $i\in \mathcal{H}$ is valid but was never produced by $i$. The view of the adversary in this hybrid is indistinguishable from the view in the previous hybrid because the signature scheme satisfies EUF-CMA security.

    \item[\textbf{Hyb$_{10}$.}] In this hybrid, Sim replaces the encryption key $sk_{i,3}$ in Step (5) with Sim($sk_{i,3}$), where $sk_{i,3}$ is chosen uniformly at random for all $i \in \mathcal{H}$ (but can still be decrypted with $pk_{i,3}$ in Step (6)). The view of $\mathcal{A}$ in this hybrid is indistinguishable from the view in the previous hybrid because of the IND-CPA security of the AE scheme and the DL assumption.

    \item[\textbf{Hyb$_{11}$.}] In this hybrid, Sim aborts if the adversary queries the random oracle on input $msk$ for $i\in \mathcal{H}$ before the shares are revealed in Step (6). Since $\mathcal{A}$ knows less than $\kappa$ shares of the master secret key $msk$. Due to the information-theoretic security of Shamir's secret sharing scheme, the view of the adversary in this hybrid is indistinguishable from the view in the previous hybrid.
    
    \item[\textbf{Hyb$_{12}$.}] In this hybrid, Sim aborts if the adversary queries the random oracle on input $g_t^{seed_i}$ for $i\in \mathcal{H}$ before the shares are revealed in Step (6). Due to the same reason as the previous hybrid, the view of $\mathcal{A}$ in this hybrid is indistinguishable from the view in the previous hybrid.

    \item[\textbf{Hyb$_{13}$.}] In this hybrid, Sim samples the values $\mathbf{y_i}$ at random in Step (3), and will program the random oracle on $seed_i$ as follows in Step (6): 
    \begin{equation*}
        \textit{PRG}(g_t^{seed_i})=\mathbf{y_i}-\mathbf{x_i}-\sum_{0<j<i}\mathbf{m_{i,j}}+\sum_{i<j}\mathbf{m_{i,j}}.
    \end{equation*}
     Since $\mathcal{A}$ has not queried the random oracle on $g_t^{seed_i}$ before Step (6), its view in this hybrid is indistinguishable from its view in the previous hybrid.

     \item[\textbf{Hyb$_{14}$.}] In this hybrid, Sim reprograms the random oracle on $seed_i$ in Step (6) as 
     \begin{equation*}
        \textit{PRG}(g_t^{seed_i})=\mathbf{y_i}-\mathbf{w_i}-\sum_{0<j<i}\mathbf{m_{i,j}}+\sum_{i<j}\mathbf{m_{i,j}},
    \end{equation*}
    subject to: $\sum_{i\in \mathcal{U_S}}\mathbf{x_i}=\sum_{i\in \mathcal{U_S}}\mathbf{w_i}$. Due to the information-theoretic security of Shamir's secret sharing scheme, the value of $\{g_t^{seed_i^{(u)}}\}_{i\in \mathcal{U_S}}$ can only be constructed when at least $\kappa$ shares have been revealed to $\mathcal{A}$, where $2\kappa> (1+\eta_{\mathcal{C}}-\eta_{\mathcal{D}})n_I$. 

    The interpretation of the threshold $\kappa$ is as follows. According to the threat model, honest decryptors will only commit to a single online and offline set and transmit corresponding secret shares of seeds. Moreover, the malicious server and corrupted decryptors cannot forge signatures on behalf of honest decryptors due to the EUF-CMA security of DS. If there were two different sets that have been committed by $\kappa$ clients, it implies that at least $\kappa-\eta_\mathcal{C}n_I$ signed both of the sets, i.e., $2(\kappa-\eta_\mathcal{C}n_I)\leq (1-\eta_{\mathcal{C}}-\eta_{\mathcal{D}})n_I$, which contradicts $2\kappa> (1+\eta_{\mathcal{C}}-\eta_{\mathcal{D}})n_I$.

    In this case, less than $\kappa$ shares of $\{g_t^{seed_{i,j}^{(u)}}\}_{j\in \mathcal{\mathcal{U_D}}}$ will be revealed to the adversary $\mathcal{A}$ since Sim will not reveal both of the shares simultaneously. Therefore, by the information-theoretic security of Shamir's secret sharing scheme, $\mathcal{A}$ cannot learn the values of $g_t^{seed_{i,j}}$ for $i,j\in \mathcal{U_S}$. By Lemma 1, the view of the adversary $\mathcal{A}$ in this hybrid is indistinguishable from the view in the previous hybrid.

    \item[\textbf{Hyb$_{15}$.}] In this hybrid, Sim programs the random oracle as above, but with random $\mathbf{w_i}$ (i.e., it does not need to condition on $\sum_{i\in \mathcal{U_S}}\mathbf{x_i} =\sum_{i\in \mathcal{U_S}}\mathbf{w_i}$). As above, since $\mathcal{A}$ cannot obtain at least $\kappa$ shares of $g_t^{seed_i}$ and $g_t^{seed_{i,j}}$ simultaneously, it cannot know the value of $g_t^{seed_i}$ and $g_t^{seed_{i,j}}$. Therefore, the view of the adversary $\mathcal{A}$ in this hybrid is indistinguishable from the view in the previous hybrid.

    \item[\textbf{Hyb$_{16}$.}] This hybrid is defined as the previous one, with the only difference being that Sim now does not receive the inputs of the honest parties, but instead, in Step (6), Sim makes one query to the functionality $\mathcal{F}_{\mathbf{x},T,\mathcal{N},\eta,\kappa}^{Fluent}$ for the set $\mathcal{U_S}\backslash \mathcal{U_C}$ and use the value to sample the required $\mathbf{w_i}$ values.

    It is easy to see that this change does not modify the view seen by the adversary $\mathcal{A}$, and therefore it is perfectly indistinguishable from the previous one. Moreover, this hybrid does not make use of the honest party’s inputs, and this concludes the proof. \hfill   $\square$
\end{enumerate}
\section{Non-interactive Threshold Signature Scheme.} \label{threshold signature scheme}

A non-interactive $(\kappa,n)_{1\leq \kappa \leq n}$ threshold signature scheme $TSS$ is a tuple of algorithms that involves $n$ nodes, and up to $\kappa-1$ nodes can be corrupted. The threshold signature scheme has the following algorithms:
\begin{itemize}[leftmargin=*]
    \item TSS.KeyGen$(1^\lambda, \kappa,n)\rightarrow \{pk,pk_i,sk_i\}.$ Given a security parameter $\lambda$, a threshold $\kappa$, and the number of shares to be generated $n$, the key generation algorithm outputs a master public key $pk$ and a vector of secret and public key shares $(sk_i,pk_i)_{i\in [n]}$.
    \item TSS.Sign$(sk_i,m) \rightarrow \sigma_i$. Given a secret key share $sk_i$ and a message $m\in\{0,1\}^*$ to be signed, the signing algorithm outputs a signature share $\sigma_i$.
    \item TSS.ShareVerify$(m,(i,\sigma_i))\rightarrow 0/1$. Given a message $m$, an index $i$ and a signature share $\sigma_i$, the deterministic share verification algorithm outputs 1 if $\sigma_i$ is a valid signature share and 0 otherwise.
    \item TSS.Combine$(m, \{(i, \sigma_i)\}_{i\in\mathcal{I}}) \rightarrow \sigma\, / \perp$. Given a message $m$, and a list of pairs $\{(i, \sigma_i)\}_{i\in \mathcal{I}}$, where $\mathcal{I}\subseteq  [n]$ and $\lvert \mathcal{I}\rvert \geq \kappa$, the deterministic combining algorithm outputs either a complete signature $\sigma$ for message $m$ or $\perp$ if an ill-formed signature share exists.
    \item TSS.Verify$(m, \sigma) \rightarrow 0/1$. Given a message $m$ and a signature $\sigma$, the deterministic signature verification algorithm outputs 1 if $\sigma$ is a valid signature and 0 otherwise.
\end{itemize}

A threshold signature scheme satisfies the following properties:
\begin{itemize}[leftmargin=*]
    \item Correctness: The property requires that $\forall m$, $\forall i\in [n]$, and $\forall \mathcal{I}\subseteq [n]$ with $\lvert \mathcal{I} \rvert \geq \kappa$, $\mathrm{Pr}[\mathrm{TSS.ShareVerify}(m,(i,\mathrm{TSS.Sign}(sk_i,m)))=1]=1$, and 
    $\mathrm{Pr}[\mathrm{TSS.Verify}(m,\mathrm{TSS.Combine}(m,\{i,\sigma_i\}_{i\in \mathcal{I}}))=1 \mid \mathrm{TSS.ShareVerify}(m,(i,\sigma_i))=1]=1$.
    \item Unforgeability: No polynomial-time adversary can forge a signature that can be verified correctly (by honest parties) of any message $m$ without querying the signature algorithm;
    \item Robustness: When a message $m$ is provided as the input of the TSS.Sign algorithm, eventually all honest parties can get a signature of $m$ that can be correctly verified.
\end{itemize}

\section{More Experimental Results} \label{appendix: more results}
A detailed experimental result in terms of different client sizes, committee sizes, and dropout rates is given in Table~\ref{table: Computational cost comparison}, where all data are averaged for a single aggregation.

\renewcommand\arraystretch{1.3}
\begin{table*}[htbp]
\caption{The concrete computational and communication costs in specific steps in various schemes, where the length of the vector are set 1.6e4.}
\begin{center}
\resizebox*{\linewidth}{!}{
\begin{tabular}{cccccccc}
\hline
Scheme & \makecell{Size of\\client} & \makecell{size of\\committee} & $\eta$ & \makecell{computational cost\\ of server (s)} & \makecell{computational cost\\ of client (s)} & \makecell{communication\\ of server (KB)} & \makecell{communication\\ of client (KB)}  \\ 
\hline

\cite{BellCCS20}  &  100 &  -- &  5\%  & 0.16 & 0.04 & 17188.96 & 172.13 \\
\cite{BellCCS20}  &  200 &  -- &  5\%  & 0.51 & 0.05 & 37153.47 & 186.00 \\
\cite{BellCCS20}  &  300 &  -- &  5\%  & 0.70 & 0.05 & 58854.08 & 196.46 \\
\cite{BellCCS20}  &  400 &  -- &  5\%  & 0.99 & 0.05 & 81392.42 &  203.77\\
\cite{BellCCS20}  &  500 &  -- &  5\%  & 1.29 & 0.05 & 105530.81 & 211.35 \\
Flamingo \cite{SP2023flamingo}  &  100 &  40 &  5\%  & 2.42 & 0.06 / 0.13 & 16792.10 & 137.69 / 84.93 \\
Flamingo \cite{SP2023flamingo}  &  200 &  40 &  5\%  & 2.99 & 0.07 / 0.25 & 33641.50 & 138.93 / 165.21 \\
Flamingo \cite{SP2023flamingo}  &  300 &  40 &  5\%  & 3.61 & 0.07 / 0.37 & 49229.00 & 137.84 / 224.36 \\
Flamingo \cite{SP2023flamingo}  &  400 &  40 &  5\%  & 4.22 & 0.08 / 0.52 & 68538.19 & 140.01 / 350.45 \\
Flamingo \cite{SP2023flamingo}  &  500 &  40 &  5\%  & 5.56 & 0.08 / 0.75 & 89460.55 & 142.14 / 506.34 \\
Fluent  &  100 &  40 &  5\%  & 1.69 & 0.01 / 0.22 & 15914.36 & 127.11/ 88.23 \\
Fluent  &  200 &  40 &  5\%  & 2.99 & 0.01 / 0.43 & 30486.01 & 127.11/ 142.55 \\
Fluent  &  300 &  40 &  5\%  & 3.58 & 0.01 / 0.57 & 44595.52 & 127.11/ 185.69 \\
Fluent  &  400 &  40 &  5\%  & 5.05 & 0.01 / 0.80 & 60801.57 & 127.11/ 280.81 \\
Fluent  &  500 &  40 &  5\%  & 6.51 & 0.01 / 1.00 & 75816.97 & 127.11/ 346.80 \\
\hline
Flamingo \cite{SP2023flamingo}  &  500 &  30 &  5\%  & 4.71 & 0.07 / 0.75 & 83978.45 & 141.15 / 506.30 \\
Flamingo \cite{SP2023flamingo}  &  500 &  35 &  5\%  & 5.56 & 0.08 / 0.76 & 86980.09 & 141.65 / 514.66 \\
Flamingo \cite{SP2023flamingo}  &  500 &  45 &  5\%  & 6.14 & 0.09 / 0.77 & 92577.38 & 142.63 / 513.74 \\
Flamingo \cite{SP2023flamingo}  &  500 &  50 &  5\%  & 6.40 & 0.09 / 0.76 & 95028.86 & 143.12 / 508.67 \\
Fluent  &  500 &  30 &  5\%  & 4.94 & 0.01 / 1.00 & 72224.02 & 127.11/ 341.77 \\
Fluent  &  500 &  35 &  5\%  & 5.51 & 0.01 / 1.00 & 73986.60 & 127.11/ 343.55 \\
Fluent  &  500 &  45 &  5\%  & 7.53 & 0.01 / 1.01 & 77646.74 & 127.11/ 348.93 \\
Fluent  &  500 &  50 &  5\%  & 8.13 & 0.01 / 1.01 & 79459.83 & 127.11/ 350.30 \\
\hline
\cite{BellCCS20}  &  500 &  -- &  <1\%  & 1.09 & 0.05 & 108082.91 & 216.43 \\
\cite{BellCCS20}  &  500 &  -- &  10\%  & 1.48 & 0.04 & 103492.85 & 207.00 \\
\cite{BellCCS20}  &  500 &  -- &  15\%  & 1.64 & 0.04 & 100436.73 & 201.12 \\
\cite{BellCCS20}  &  500 &  -- &  20\%  & 1.83 & 0.04 & 98887.36 & 197.99 \\
Flamingo \cite{SP2023flamingo}  &  500 &  40 &  <1\%  & 2.68 & 0.08 / 0.19 & 79531.19 & 142.93 / 204.82 \\
Flamingo \cite{SP2023flamingo}  &  500 &  40 &  10\%  & 8.07 & 0.08 / 1.30 & 99031.34 & 141.34 / 797.15 \\
Flamingo \cite{SP2023flamingo}  &  500 &  40 &  15\%  & 10.13 & 0.08 / 1.77 & 107278.89 & 140.51 / 1059.03 \\
Flamingo \cite{SP2023flamingo}  &  500 &  40 &  20\%  & 11.74 & 0.08 / 2.16 & 113965.27 & 139.74 / 1274.46 \\

Fluent  &  500 &  40 &  <1\%  & 4.25 & 0.01 / 0.83 & 75010.47 & 127.11/ 286.51 \\
Fluent  &  500 &  40 &  10\%  & 8.74 & 0.01 / 1.18 & 76595.91 & 127.09/ 405.72 \\
Fluent  &  500 &  40 &  15\%  & 11.05 & 0.01 / 1.37 & 77368.49 & 127.07/ 464.91 \\
Fluent  &  500 &  40 &  20\%  & 13.44 & 0.01 / 1.56 & 78380.86 & 127.06/ 529.77 \\

\hline
\end{tabular}
}
\label{table: Computational cost comparison}
\end{center}
\end{table*}

\end{document}